\documentclass[aps,prx,twocolumn,groupedaddress,showpacs,preprintnumbers]{revtex4-1}
\usepackage{epsfig}
\usepackage{graphics}
\usepackage{color}
\usepackage{calc}
\usepackage{hyperref}
\usepackage{array}
\usepackage{mathtools}
\usepackage{graphicx}
\usepackage[bitstream-charter, greekfamily=default]{mathdesign}
\usepackage{lipsum}
\usepackage{comment}

\usepackage{xcolor}

\newcommand{\de}{\mathrm{d}}
\newcommand{\im}{\mathrm{Im}}
\newcommand{\De}{\mathcal{D}}
\newcommand{\avg}[1]{\left\langle #1 \right\rangle}
\newcommand{\abs}[1]{\left\vert #1 \right\vert}
\newcommand{\mt}[1]{\textbf{{\color{teal}[#1]}}}
\newcommand{\eg}{{\it e.g.}}
\newcommand{\ie}{{\it i.e.}}

\usepackage{physics}
\usepackage[capitalize]{cleveref}

\begin{document}

\title{The localized phase of the Anderson model on the Bethe lattice}

\author{Tommaso Rizzo\textsuperscript{1,2} and Marco Tarzia\textsuperscript{3,4}}

\affiliation{\textsuperscript{1}\mbox{Dip. Fisica, Universit\`a ``Sapienza'', Piazzale A. Moro 2, I–00185, Rome, Italy}\\
	\textsuperscript{2}\mbox{ISC-CNR, UOS Rome, Universit\`a ``Sapienza'', Piazzale A. Moro 2, I-00185, Rome, Italy}\\
	\textsuperscript{3} \mbox{LPTMC, CNRS-UMR 7600, Sorbonne Universit\'e, 4 Pl. Jussieu, F-75005 Paris, France}\\
	\textsuperscript{4} \mbox{Institut  Universitaire  de  France,  1  rue  Descartes,  75231  Paris  Cedex  05,  France}}

\begin{abstract}
In this paper, we investigate the Anderson model on the Bethe lattice, focusing on the localized regime. Employing the cavity approach, we derive compact expressions for the inverse participation ratios (IPRs) that are equivalent to those obtained using the supersymmetric formalism and naturally facilitate a highly efficient computational scheme. This method yields numerical results with unprecedented accuracy, even very close to the localization threshold. Our approach allows for high-precision validation of all theoretical predictions from the analytical solution, including the finite jump of the IPRs at the transition. Additionally, we reveal a singular behavior of the IPRs near the critical point that has not been previously reported in the literature. This singular behavior is further confirmed by the numerical solution of the non-linear $\sigma$ model on the Bethe lattice, which provides an effective description of Anderson localization.
\end{abstract}

\maketitle 

\section{Introduction}

Anderson localization (AL)~\cite{anderson1958absence} is one of the most spectacular phenomenon in condensed matter physics. It manifests as the suppression of wave propagation in a disordered medium above a critical value of the disorder strength (and for any finite disorder in low enough dimension)~\cite{lee1985disordered,evers2008anderson,lagendijk2009fifty}. Over the past half-century the field has thrived, with recent experimental observations in diverse systems such as 
cold atomic gases~\cite{aspect2009anderson,roati2008anderson,billy2008direct,kondov2011three,jendrzejewski2012three,semeghini2015measurement}, 
kicked rotors~\cite{chabe2008experimental}, and classical sound elastic waves~\cite{hu2008localization} further highlighting the ubiquity and relevance of this phenomenon. 

On the theoretical side, the critical properties of AL are well established in low dimensions. According to the scaling hypothesis~\cite{abrahams1979licciardello} $d_L=2$ is the lower critical dimension of the transition (for systems with orthogonal symmetry)~\cite{mott1961theory,gor1996particle}. The scaling arguments have been later supported and quantitatively confirmed by a renormalization group analysis in $d = 2 + \epsilon$~\cite{hikami1992localization,foster2009termination} of an effective field-theory description in terms of a non-linear $\sigma$ model (NL$\sigma$M)~\cite{wegner1979mobility,schafer1980disordered,efetov1983supersymmetry}.

AL is also analytically tractable in the infinite-dimensional limit on the Bethe lattice (BL)~\cite{anderson1973selfconsistent}, an infinite tree (with no boundaries) in which each node as a fixed degree $k+1$. The hierarchical structure of the BL allows one to obtain a (complicated) non-linear integral self-consistent equation for the order parameter distribution function, which becomes asymptotically exact in the thermodynamic limit, and whose  analysis yields the transition point and the critical behavior~\cite{anderson1973selfconsistent,efetov1985anderson,efetov1987density,efetov1987anderson,zirnbauer1986localization,zirnbauer1986anderson,verbaarschot1988graded,mirlin1991localization,mirlin1991universality,fyodorov1991localization,fyodorov1992novel,mirlin1994statistical,tikhonov2019statistics,tikhonov2019critical,biroli2010anderson,parisi2019anderson,pascazio2023anderson}. 

Despite these results have been firmly established already several years ago, the study of AL on the BL is still very active, and has continued to reveal new facets and intricacies. There are two main reasons for this. The first concerns the differences between the exotic critical behavior found on the BL and the one observed in finite dimensions and predicted by the scaling analysis. In particular,   the diffusion coefficient (or the conductivity) 
vanishes exponentially at the critical disorder on the BL when the transition is approached from the metallic side~\cite{efetov1985anderson,efetov1987density,efetov1987anderson,zirnbauer1986localization,zirnbauer1986anderson,verbaarschot1988graded,mirlin1991localization,mirlin1991universality,fyodorov1991localization,fyodorov1992novel,mirlin1994statistical,tikhonov2019statistics,tikhonov2019critical,biroli2022critical}, while in finite-$d$ such exponential behavior 
is replaced by a power law (with a $d$-dependent exponent $\nu (d-2)$, $d$ being the spatial dimension and $\nu$ the critical exponent describing the divergence of the localization length at the critical disorder~\cite{mirlin1994distribution,tarquini2017critical,lee1985disordered,evers2008anderson}). The other difference concerns the behavior of the inverse participation ratio (IPR). The IPR is defined as $I_2 = \langle \sum_{i=1}^N \vert \psi_\alpha (i) \vert^4 \rangle$, 
and is essentially a measure of the inverse volume occupied by an eigenstate. On BLs of finite size $N$ (\ie, random-regular graphs in which every node has a fixed connectivity $k+1$~\cite{wormald1999models}, see below for a precise definition), $I_2 \simeq \Lambda/N$ in the metallic phase (with a disorder-dependent prefactor $\Lambda$ which 
diverges exponentially for $W \to W_c^{-}$~\cite{tikhonov2019statistics,baroni2024corrections}), exhibits a discontinuous jump at the transition, and stays of $O(1)$ for $W>W_c$~\cite{efetov1985anderson,efetov1987density,efetov1987anderson,zirnbauer1986localization,zirnbauer1986anderson,verbaarschot1988graded,mirlin1991localization,mirlin1991universality,fyodorov1991localization,fyodorov1992novel,mirlin1994statistical,tikhonov2019statistics}. In contrast, in finite-dimensional systems the IPR vanishes as a power-law at the critical disorder with an exponent $\nu d$~\cite{evers2008anderson,mirlin1994distribution}. 
Several works have addressed these apparent discrepancies. Both intuitive arguments and quantitative calculations~\cite{mirlin1994distribution,baroni2024corrections} have provided strong indications of the fact that the BL limit is a singular point of AL and plays the role of the upper critical dimension of the problem, $d_L = \infty$, in agreement with previous conjectures~\cite{tarquini2017critical,garcia2007dimensional,castellani1986upper,mard2017strong}. 

The second reason for the remarkable resurgence of interest in AL on the BL and sparse random graphs can be attributed to its strong connection with many-body localization (MBL)~\cite{basko2006metal,gornyi2005interacting}. MBL involves the localization of highly excited many-body eigenstates even in the presence of interactions, and has been a focal point of recent theoretical and experimental research~\cite{altman2015universal,nandkishore2015many,abanin2017recent,alet2018many,abanin2019colloquium}. 
Since the preliminary investigations, MBL was linked to a form of localization in the Fock space of Slater determinants~\cite{altshuler1997quasiparticle} (see also Refs.~\cite{basko2006metal,gornyi2005interacting,herre2023ergodicity}): In this representation, many-body configurations correspond to site orbitals on the graph, subject to (strongly correlated) diagonal disorder, while interactions serve as effective hoppings connecting them. Despite several simplifications in this analogy, 
it proves valuable for qualitatively understanding the problem~\cite{tikhonov2021anderson,de2013ergodicity,biroli2017delocalized,biroli2020anomalous,logan2019many,garcia2022critical,herre2023ergodicity,biroli2023large}. 

In this context, a set of analytical~\cite{tikhonov2019statistics,tikhonov2019critical,baroni2024corrections,parisi2019anderson,zirnbauer2023wegner,arenz2023wegner,biroli2010anderson,biroli2022critical,bapst2014large} and numerical~\cite{garcia2022critical,biroli2012difference,de2014anderson,altshuler2016nonergodic,kravtsov2018non,savitz2019anderson,pino2020scaling,pino2023correlated,tikhonov2016anderson,garcia2017scaling,biroli2018delocalization,bera2018return,de2020subdiffusion,vanoni2023renormalization,garcia2020two,sierant2023universality} explorations of the Anderson model on the BL 
has been conducted over the last decade. 
In the midst of such numerous investigations, the predominant research emphasis has leaned towards the delocalized side preceding the critical disorder, leaving the insulating regime relatively underexplored, with only a few notable exceptions~\cite{garcia2020two,garcia2022critical}. Bridging this gap, in this work we perform a thorough investigation of the critical properties of AL on the infinite BL when the transition is approached from the localized phase. Possibly one reason the insulating phase has received comparatively less attention may be attributed to the inherent challenge posed by the fact that the order parameter distribution function (\ie, the probability distribution of the local density of states, see below) exhibits power law tails, which are exceptionally difficult to sample accurately using conventional numerical methods. In fact, the first  achievement of our work precisely consists in circumventing this problem: Using the cavity formalism, we derive compact and transparent expressions for the relevant observables (such as the IPR and the distribution of the wave-functions' amplitudes) 
that are equivalent to those obtained within the supersymmetric approach.  These expressions lend themselves naturally to a highly efficient computational method allowing us to obtain results
with unprecedented numerical even very close to the transition point. This enables us to precisely assess and validate all the predictions of the analytical solution~\cite{anderson1973selfconsistent,efetov1985anderson,efetov1987density,efetov1987anderson,zirnbauer1986localization,zirnbauer1986anderson,verbaarschot1988graded,mirlin1991localization,mirlin1991universality,fyodorov1991localization,fyodorov1992novel,mirlin1994statistical,tikhonov2019statistics} and recover the expected critical behavior~\cite{tikhonov2019critical}. In particular our results clearly show that that the IPR exhibit a finite jump at the localization transition, as predicted by the supersymmetric treatment~\cite{efetov1985anderson,efetov1987density,efetov1987anderson,zirnbauer1986localization,zirnbauer1986anderson,verbaarschot1988graded,mirlin1991localization,mirlin1991universality,fyodorov1991localization,fyodorov1992novel,mirlin1994statistical,tikhonov2019statistics}. This is particularly interesting, as the existence of such a finite jump has been questioned in some recent works~\cite{garcia2022critical}.

The second noteworthy outcome of our investigation unveils a distinctive feature: the finite jump of the IPR at the critical point is followed by a square root singularity which, to the best of our knowledge, has never been reported in existing literature. Such singular behavior is further corroborated by the solution of the self-consistent equations found on the BL for the NL$\sigma$M which provides an effective description of AL~\cite{efetov1985anderson,efetov1987density,efetov1987anderson,zirnbauer1986localization,zirnbauer1986anderson,verbaarschot1988graded}. The analysis of the NL$\sigma$M also helps to elucidate the highly non-trivial mathematical mechanism underlying this square root singularity.

The paper is organized as follows: In Sec.~\ref{sec:model} we introduce the Anderson tight-binding model on the BL and briefly recall the definition of the key observables and the main features of its analytical solution; In Sec.~\ref{sec:critical} we discuss the linearized self-consistent equations with respect to a small imaginary part which describe the localized phase and review the main features of their critical behavior; The main results of our work are contained in Sec.~\ref{sec:IPR}: We start by presenting our new approach to accurately solve the linearized equations, which allows one to retrieve the full probability distribution of the wave-functions' amplitudes in the localized phase; We discuss the resulting singular behavior of the IPR close to the transition point; We show that the solution of the effective NL$\sigma$M fully support our findings; 
Finally, in Sec.~\ref{sec:conclusions} we provide a summary of our results and a few perspectives for future investigations. In the Appendix sections~\ref{app:Kernel}--\ref{app:ED} we  present  some  technical details and supplementary  information  that  complement the results discussed in the main text.

\section{The model and known results} \label{sec:model}

We consider the simplest  model for AL, which consists in a non-interacting (spinless) quantum particle on a lattice in presence of a disordered potential:
\begin{equation} \label{eq:H}
	H = -  \sum_{\langle i, j \rangle} t_{ij} \left( \vert i \rangle \langle j \vert + \vert j \rangle \langle i \vert \right)  - \sum_{i=1}^N \epsilon_i \vert i \rangle \langle i \vert \, .
\end{equation}
The first term is a sum over all pairs of nearest neighbors sites and corresponds to 
the adjacency matrix of the considered lattice ($t_{ij}$ is the hopping kinetic energy scale, which we take equal to 1 throughout). The second sum runs over all $N$ sites of the lattice and corresponds to a diagonal random matrix containing the disordered potential. The on-site energies $\epsilon_i$ are independent and identically distributed random variables. It is custom to extract them accordingly to a uniform distribution in the interval $[-W/2, W/2]$, $W$ being the disorder strength. The model is defined on an {\it infinite} BL, which is formally described as an infinite random-regular graph (RRG)~\cite{wormald1999models} in which each vertex has a fixed degree $k+1$, and can be thought as a tree wrapped onto itself and without boundaries. In fact it can be rigorously shown that RRGs of $N$ nodes have locally a tree-like structure and loops whose typical length scales as $\ln N/\ln k$~\cite{wormald1999models}. For concreteness in the following we mostly focus on the case $k=2$, but the same qualitative behavior is expected for any finite $k$ strictly larger than $1$.

The order parameter associated to AL is the probability distribution of the local density of states (LDoS)~\cite{mirlin1994distribution}, defined as
\begin{equation} \label{eq:LDoS}
\rho_i (E)  \equiv \sum_\alpha |\psi_\alpha (i)|^2 \delta(E-E_\alpha ) \, ,
\end{equation}
where $\psi_\alpha$ and $E_\alpha$ are the eigenvectors and the eigenvalues of $H$. Physically $\rho_i(E)$ is tightly related to the inverse lifetime of a  particle of energy $E$ created in $i$, and its typical value is proportional to the diffusion constant (or the dc conductivity). In the insulating phase the LDoS vanishes in the thermodynamnic limit for $W>W_c$, since the exponentially localized eigestates of energy $E$ are typically very far from a given node $i$ and do not contribute to the sum. Instead in the metallic phase the LDoS is finite with probability density $P(\rho)$, since extended plane waves have typically  amplitudes of order $1/N$ on all the nodes of the graph. 

Localization begins from the band edges~\cite{abou1974self,biroli2010anderson}, therefore to see if all states are localized it is sufficient to look at the band center. Hence, for simplicity, we set $E=0$ throughout the rest of the paper. 

The distribution of the LDoS, as well as other properties of the spectral statistics of the model, are encoded in the statistics of the  elements of the resolvent matrix~\cite{economou2006green}, defined as
\begin{equation} \label{eq:resolvent}
    G_{ij} = ({\rm i} \eta {\mathbb I} - H)^{-1}_{ij} \, ,
\end{equation}
where ${\mathbb I}$ is the identity matrix, $H$ is the Hamiltonian~\eqref{eq:H}, and $\eta$ is an infinitesimal imaginary regulator that softens the pole singularities in the denominator of $G$. On the BL the diagonal elements of $G$ verify a set of self-consistent recursion relation~\cite{anderson1973selfconsistent,biroli2010anderson,economou2006green}, which become asymptotically exact in the $N \to \infty$ limit~\cite{bordenave2010resolvent}. Deriving these equations involves considering the resolvent matrices of modified Hamiltonians $H^{(i)}$, where the node $i$ has been removed from the lattice (\ie, $H^{(i)}$ is obtained by eliminating the $i$-th row and column from $H$). The crucial observation here is that, owing to the hierarchical structure of the BL, removing one node renders each of its neighbors uncorrelated from the others, as the lattice breaks into $k+1$ semi-infinite disconnected branches. Consequently, on any given site $i$ one obtains (\eg, by direct Gaussian integration~\cite{Dean_2002,Rogers_2008,Susca_2021,biroli2010anderson,economou2006green} or by using the block matrix inversion formula, also called the Schur complement formula~\cite{arous2008spectrum}):
\begin{equation} \label{eq:cavity}
G_{i \to j} =\frac{1}{ \epsilon_i - {\rm i} \eta - \sum_{m \in \partial i \setminus j} t_{mi}^2 G_{m \to i} }\, ,
\end{equation}
where $G_{i \to j}=({\rm i} \eta {\mathbb I} -  H^{(j)})^{-1}_{ii}$ are the so-called ``cavity'' Green's functions (\ie, the diagonal element on node $i$ of the resolvent of the Hamiltonian $H^{(j)}$ obtained by removing the node $j$), $\epsilon_i$ is the on-site random energy taken from the
uniform distribution, and $\partial i \setminus j$ denotes the set of all $k + 1$ neighbors of $i$ except $j$. (Note that for each node one can define $k + 1$ cavity Green’s functions, each one satisfying a recursion relations of this kind when one of the $k+1$ neighbors of the node has been removed.) From the solution of these equations one can finally obtain the diagonal elements of the resolvent on the node $i$ of the original problem as:
\begin{equation} \label{eq:green}
G_{ii} =\frac{1}{ \epsilon_i - {\rm i} \eta -  \sum_{m \in \partial i} t_{mi}^2 G_{m \to i} }\, .
\end{equation}
Eq.~(\ref{eq:cavity}) should be in fact interpreted as a self-consistent integral equation for the probability distribution of the cavity Green's functions (in the $N \to \infty$ limit)
\begin{equation}
\begin{aligned}
P(\Re G, \Im G) =& 
\bigg \langle \frac{1}{N(k+1)} \sum_{i=1}^N \sum_{j \in \partial i} \delta(\Re G - \Re G_{i \to j}) \times \\
& \qquad \;\;\;\;\; \times \delta(\Im G - \Im G_{i \to j}) \bigg \rangle \, ,
\end{aligned}
\end{equation}
where the average is performed over the disorder distribution.
Such integral self-consistent equation can be solved numerically using population dynamics algorithms: The probability distribution of the cavity Green's functions is approximated by the empirical distribution of a large pool of $\Omega$ complex elements $(\Re G_\alpha, \Im G_\alpha)$, $P(\Re G,\Im G) \simeq \omega^{-1} \sum_{\alpha=1}^\Omega \delta (\Re G - \Re G_\alpha) \delta (\Im G - \Im G_\alpha)$; At each iteration step $k$ instances  $(\Re G_\alpha, \Im G_\alpha)$ are extracted from the pool and a value of $\epsilon$ is taken at random from the uniform distribution; A new instance of $(\Re G_\alpha, \Im G_\alpha)$ is generated using Eq.~\eqref{eq:cavity} and inserted in a random position of the pool until the process converges to a stationary distribution (convergence can be monitored for instance by checking that some moments of $P(\Re G,\Im G)$ reach a stationary value). Once the fixed stationary distribution of the cavity Green's function is found, one can implement a similar procedure to obtain the probability distribution ${\cal P}(\Re G, \Im G)$ of the Green's function of the original problem from Eq.~\eqref{eq:green} (see also Refs.~\cite{mezard2001bethe,biroli2022critical,biroli2018delocalization,tikhonov2019critical} for more details).

It is easy to show that the LDoS on a given node $i$ of the lattice, Eq.~\eqref{eq:LDoS}, is proportional to the imaginary part of the Green's function (in the $\eta \to 0^+$ limit):
\begin{equation}
\label{eq:ldos} \rho_i = \frac{1}{\pi} \lim_{\eta \to 0^+} {\rm Im} G_{ii} \, ,
\end{equation}
from which the average density of states (DoS) at $E=0$ is simply given by 
\begin{equation} \label{eq:DoS}
    \avg{\rho} = \frac{1}{N} \sum_i \delta (E_\alpha) = \frac{1}{N} \sum_i \rho_i = \frac{1}{\pi} \avg{ \Im G_{ii}} \, .
\end{equation}
Similarly, the generalized inverse participation ratios, defined as
\begin{equation} \label{eq:Ipdef}
    I_p = \frac{\left \langle \sum_\alpha \sum_i |\psi_\alpha (i)|^{2 p} \delta (E_\alpha) \right \rangle }
    {\left \langle \sum_\alpha \delta (E_\alpha) \right \rangle }\, , 
\end{equation}
are associated to the $p$-th moments of the Green's functions in the limit $\eta \rightarrow 0^+$: 
\begin{equation} \label{eq:giip}
\begin{aligned}
|G_{ii}|^p = & \left|\sum_{\alpha} |\psi_\alpha (i) |^2 \frac{1}{{\rm i} \eta - E_\alpha} \right|^p 
 \approx \sum_{\alpha} |\psi_\alpha (i)|^{2 p} \frac{1}{(\eta^2+E_\alpha^2)^{p/2}} \\
= & \sum_{\alpha} |\psi_\alpha (i)|^{2 p} \delta(E_\alpha) \frac{1}{\eta^{p-1}}\int _{-\infty}^{+\infty}\frac{1}{(1 + x^2)^{p/2}}   \, \de x \, .
\end{aligned}
\end{equation}
Averaging over all sites, from Eqs.~\eqref{eq:DoS},~\eqref{eq:Ipdef}, and~\eqref{eq:giip} one obtains a simple spectral representation of the generalized IPRs (for $p>1$):
\begin{equation} \label{eq:Ip}
    I_p = \frac{\sqrt{\pi} \, \Gamma \left ( \frac{p}{2} \right)}{\Gamma \left( \frac{p-1}{2} \right)} \lim_{\eta \to 0^+} \frac{\eta^{p-1} \left \langle |G_{ii}|^p \right \rangle}{\left \langle {\rm Im} G_{ii} \right \rangle } \, .
\end{equation}

In the metallic phase $P(\Re G, \Im G)$ (and consequently ${\cal P}(\Re G, \Im G)$) converges to a stable non-singular distribution. 
Hence, $\left \langle |G|^p \right \rangle$ is finite and from Eq.~\eqref{eq:Ip} one immediately sees that all the generalized IPR vanish for $\eta \to 0^+$. 

In the insulating phase, instead, $P(\Re G, \Im G)$ (and consequently ${\cal P}(\Re G, \Im G)$) is singular in the $\eta \to 0$ limit: The (marginal) probability distribution of $\Im G$ has a maximum in the region ${\rm Im}G \sim \eta$ and power-law tails $P({\rm Im} G) \sim \sqrt{\eta}/({\rm Im} G)^{3/2}$ with a cutoff at $\eta^{-1}$. Hence the main contribution to the moments comes from the cutoff, $\langle ({\rm Im} G)^p \rangle \propto \eta^{1-p}$ (for $p \ge 1/2$). The generalized IPR are all of $O(1)$ for $W \ge W_c$, and have a finite jump at the transition. (The average DoS, instead, is continuous across the transition.) The normalization integral is dominated by the region ${\rm Im} G \sim \eta$, and the typical value of LDoS is of order $\eta$. This behavior reflects the fact that in the localized phase wave-functions are exponentially localized on few $O(1)$ sites where $\rho_i$ takes very large values, while the typical value of the LDoS is exponentially small and vanishes in the thermodynamic limit for $\eta \to 0^+$.

\section{The self-consistent equations in the linearized regime and the critical behavior} \label{sec:critical}

As discussed above, in the localized phase the imaginary part of the Green's functions goes to zero linearly with $\eta$. It is then convenient to write in full generality
\[
    G_{i \to j} = g_{i \to j}+ \eta \, {\rm i} \,  \hat{g}_{i \to j} \ .
\]
When $\eta$ is small the cavity equation can be linearized and can be rewritten as (we set $t_{ij}=1$ on all edges $\langle i,j \rangle$ throughout):
\begin{eqnarray}  
{g}_{i \to j} &=& \frac{1}{\epsilon_i - \sum_{m \in \partial i \setminus j} \, {g}_{m \to i} }\, , \label{eq:real} \\
\hat{g}_{i \to j} &= &{g}_{i \to j}^2\left( 1+  \sum_{m \in \partial i \setminus j} \hat{g}_{m \to i} \right)\, . \label{eq:imag}
\end{eqnarray}
The above equations have two important features: i) The equation for the real part does not depend on the imaginary part, as the ${g}_{i \to j}$'s obey the equation corresponding to $\eta=0$; ii) The equation for the imaginary part is linear and thus it does not depend on $\eta$.

\begin{figure}
\includegraphics[width=0.49\textwidth]{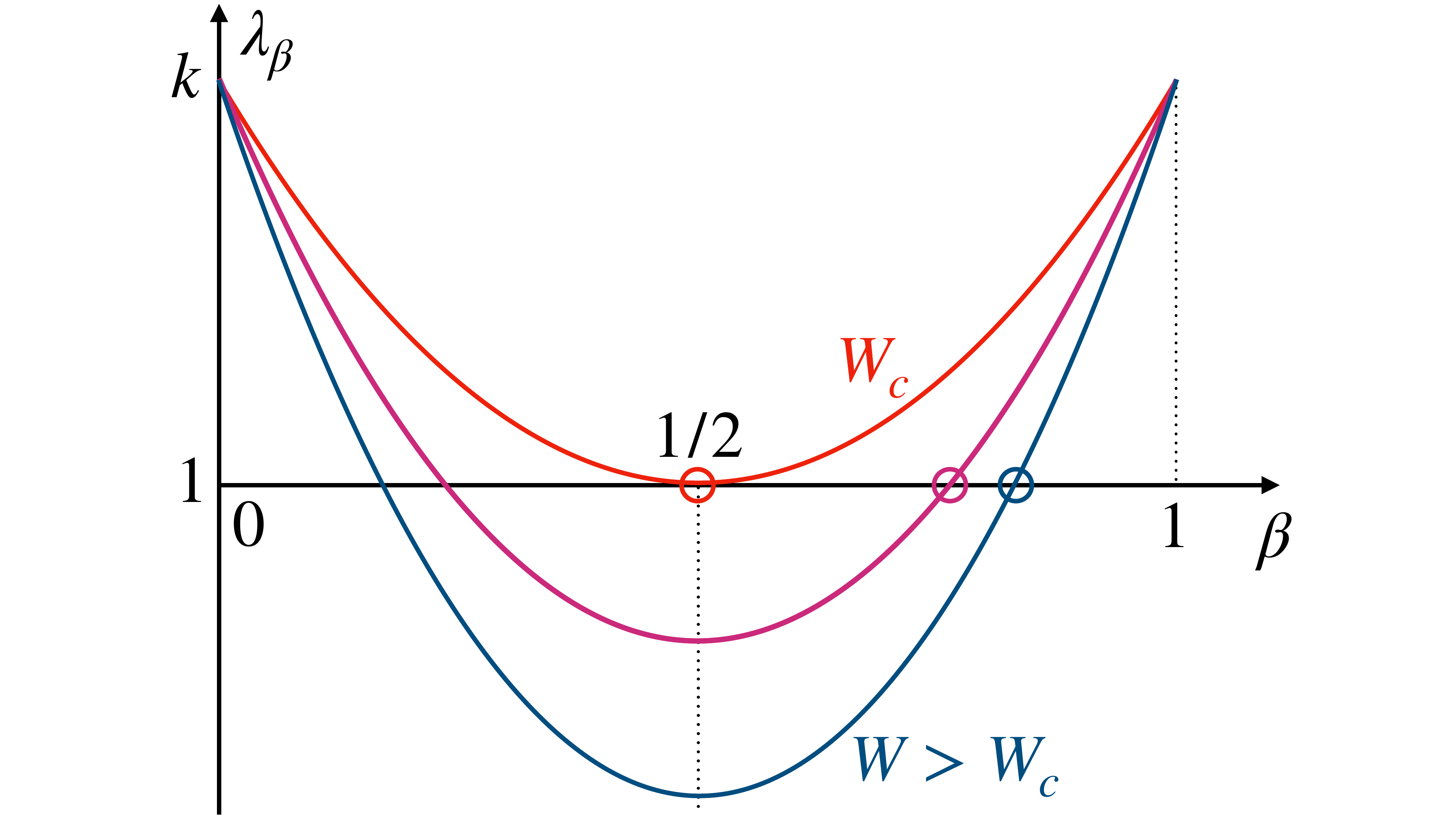} 
\caption{Illustration of the behavior of the largest eigenvalue of the integral operator~\eqref{eq:Kernel} as a function of $\beta \in [0,1]$ for $W \ge W_c$
\label{fig:kernel}}
\end{figure}

The critical disorder $W_c$ is found by studying the stability of the linearized equations~\eqref{eq:real} and~\eqref{eq:imag}~\cite{tikhonov2019critical,parisi2019anderson,anderson1973selfconsistent,tarquini2016level,bapst2014large,mirlin1991localization}. After some manipulations (see App.~\ref{app:Kernel}) one finds that a solution of the linearized equations of the form
\begin{equation} \label{eq:ansatz}
P(g,\hat{g}) \simeq \frac{f(g)}{\hat{g^{1 + \beta}}} \qquad \qquad \textrm{(for $\hat{g} \gg 1$)}
\end{equation}
only exists if the function $f(g)$ satisfies the following integral equation:
\begin{equation} \label{eq:LIO}
f(g) = \int K_\beta (g,g_1) f(g_1) \, \de g_1 \, ,
\end{equation}
which defines a linear $\beta$-dependent integral operator with the (non-symmetric) kernel~\cite{tikhonov2019critical,parisi2019anderson,anderson1973selfconsistent,mirlin1991localization} 
\begin{equation} \label{eq:Kernel}
K_\beta (g,g_1) = k \, |g|^{2 \beta} \!\! \int \!\! \de \epsilon p(\epsilon) \, \de \tilde{g}  \tilde{P} (\tilde{g}) \, \delta \left( g - \frac{1}{\epsilon - g_1 - \tilde{g}} \right)  \, . 
\end{equation}
Here $p (\epsilon)$ is the uniform box distribution of width $W$ of the random energies and $\tilde{P} (\tilde{g})$ is the probability distribution of the sum of the real part of $k-1$ cavity Green’s functions, Eq.~\eqref{eq:Ptilde}.
In order for the localized phase to be stable the largest eigenvalue $\lambda_\beta$ of the integral operator must be smaller than 1. 
It is possible to show (see App.~\ref{app:Kernel}) that, due to a symmetry of the problem~\cite{tikhonov2019critical,parisi2019anderson,anderson1973selfconsistent,mirlin1991localization}, for each left eigenvector $\phi(g)$ of the integral operator 
the function $|g_1|^{-2 \beta} \phi(1/g_1)$ is also a right eigenvector (with the same eigenvalue) of the integral operator with $\beta \to 1 - \beta$. Hence the spectrum of~\eqref{eq:Kernel}, and in particular its largest eigenvalue, must be symmetric around $\beta=1/2$, as schematically illustrated in Fig.~\ref{fig:kernel}. The condition that Eq.~\eqref{eq:LIO} admits a solution fixes the value of $\beta$ (the solution with $\beta>1/2$ must be picked since in the strong disorder limit one has that $\beta \to 1$). The critical point is identified by the point where the solution no longer exists (\ie, the largest eigenvalue of the integral operator becomes larger than one for any $\beta$). Due to the symmetry $\beta \to 1 - \beta$, at the transition point one has that $\beta=1/2$\cite{anderson1973selfconsistent}. 

One can estimate the largest eigenvalue numerically by suitably discretizing the Kernel~\eqref{eq:Kernel} on a finite grid. This has been recently been done for $k=2$ with great accuracy in Ref.~\cite{tikhonov2019critical} (see also Ref.~\cite{parisi2019anderson}). One finds that the largest eigenvalue for $W$ close to $W_c$ and for $\beta$ close to $1/2$ behaves as:
\begin{equation} \label{eq:lambda_beta}
    \lambda_\beta \simeq 1 - c_1 (W-W_c) + c_2 \left(\beta -\frac{1}{2} \right)^2 \, ,
\end{equation}
with the numerical coefficients given in Ref.~\cite{tikhonov2019critical}:
\begin{equation} \label{eq:coefficients}
W_c \simeq 18.17 \, , \qquad c_1 \simeq 0.0308 \, , \qquad c_2 \simeq 3.18 \, .
\end{equation}
For $W \gtrsim W_c$ we thus have
\begin{equation}
    \label{eq:beta}
    \beta \simeq \frac{1}{2} + \sqrt{\frac{c_1}{c_2}}  \sqrt{W - W_c}\, .
\end{equation}

\begin{figure*}
\includegraphics[width=0.42\textwidth]{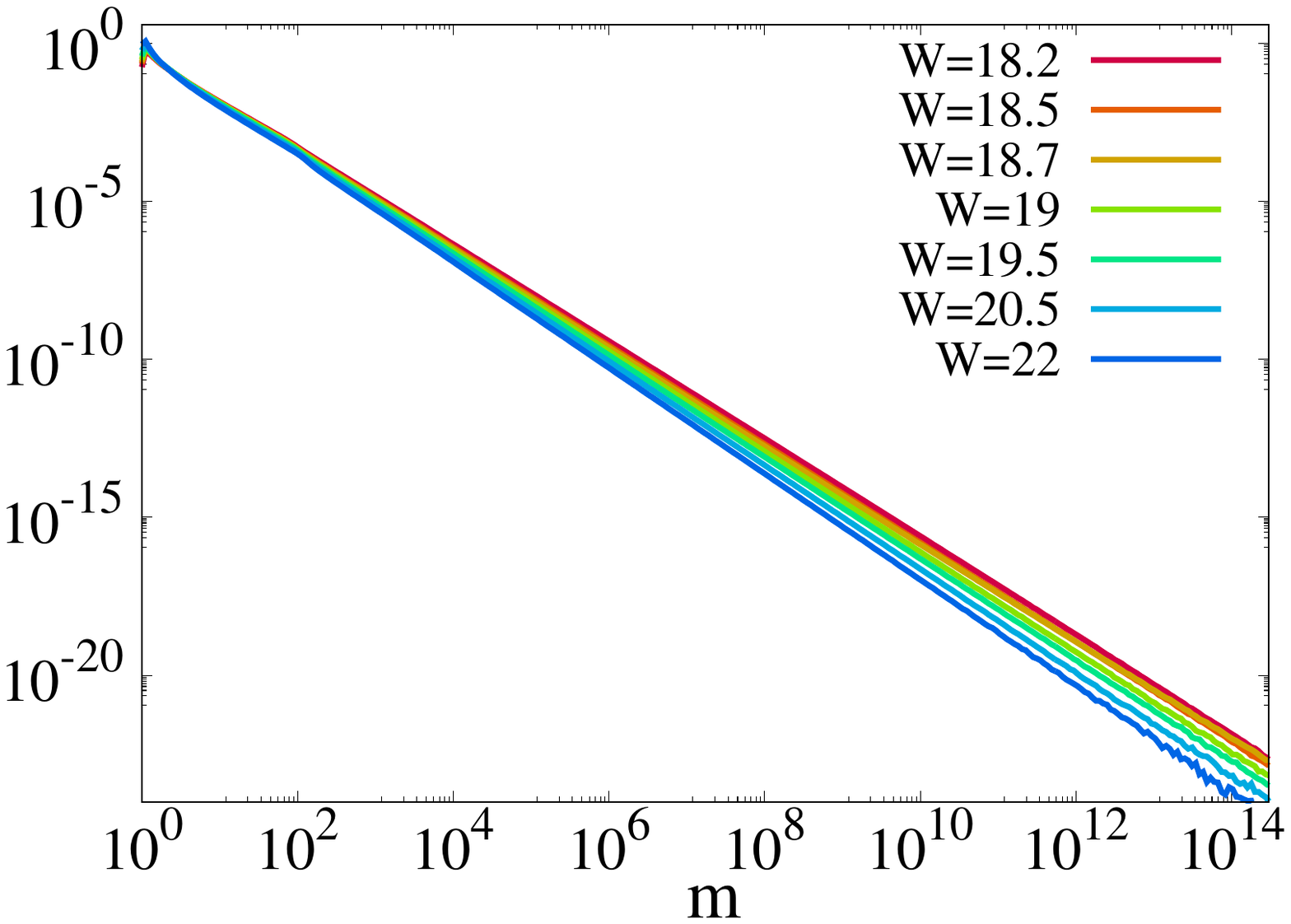} 
\put(-215,74){\rotatebox[origin=c]{90}{\large $Q(0,\hat{m})/\avg{\rho}$}} \put(-96,2){\large \^}\hspace{0.5cm}
\includegraphics[width=0.42\textwidth]{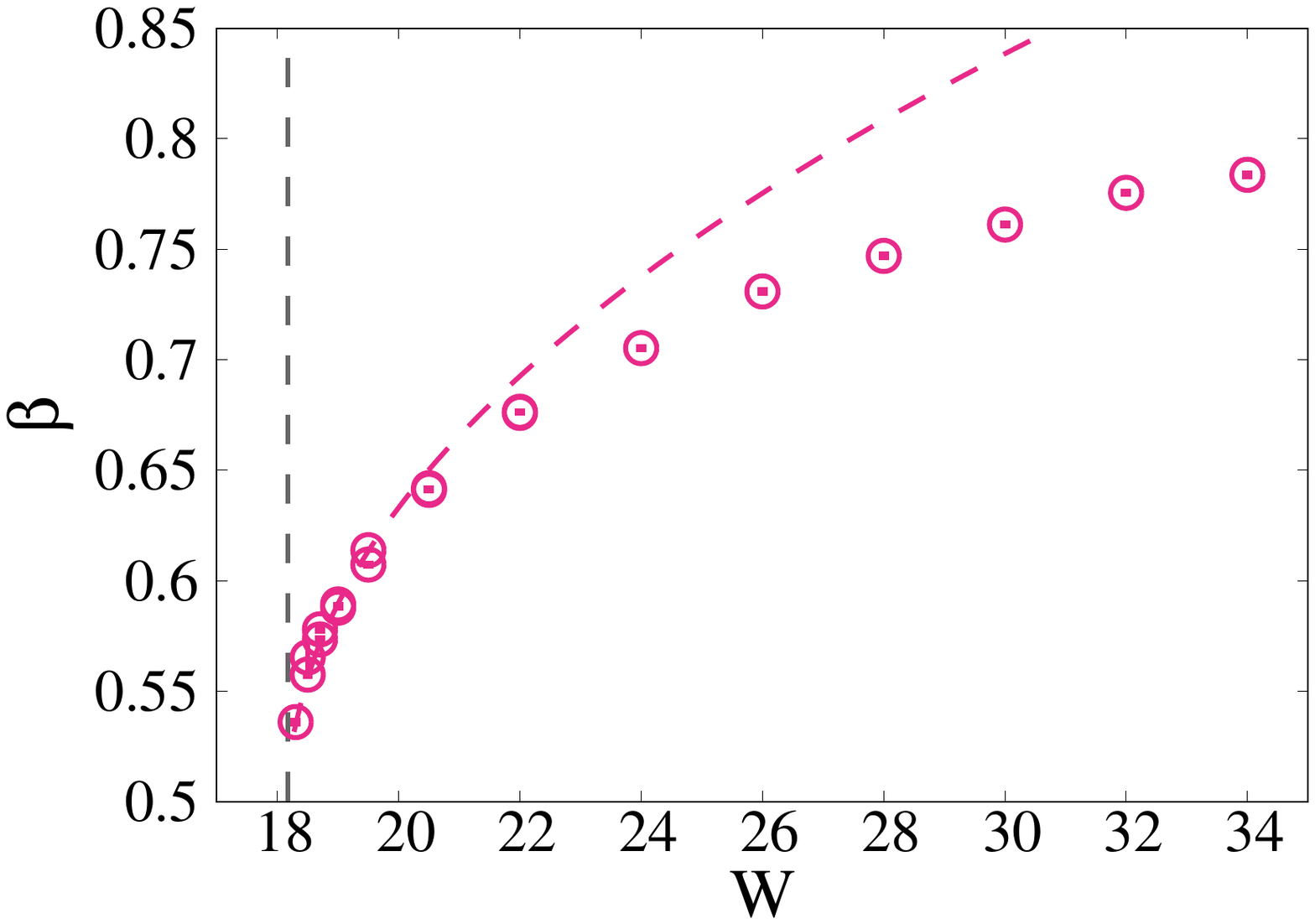} 
\caption{Left: Probability distribution $Q(0,\hat{m})/\avg{\rho}$ 
obtained setting the real part $m$ to zero using the procedure described in the text. 
Right: Exponent $\beta$ of the power-law tails of the probability distributions as a function of $W$. The dashed curve shows the prediction of Eq.~\eqref{eq:beta}  close to $W_c$ obtained  from Ref.~\cite{tikhonov2019critical}, which is in excellent agreement with our estimation. (The symbols' size exceeds that of the error bars.)
\label{fig:Pmhat}}
\end{figure*}

\section{Critical behavior of the Inverse Participation Ratio} \label{sec:IPR}

In this section we introduce suitable variables whose typical values are related to the (generalized) IPR's, and describe an algorithm that allows one to compute the $I_p$'s  with very high numerical accuracy, arbitrarily close to the critical point.

The generalized IPR's are related via Eq.~\eqref{eq:Ip} to the $p$-th moment of $|G_{ii}|$  which is broadly distributed, according to Eq.~\eqref{eq:ansatz}. The power-law tails of its probability distribution would lead to divergent expressions for $\langle |G_{ii}|^p \rangle$. In practice this does not occur because the power-law behavior is cut off at large values of the imaginary part at $\eta^{-1}$, which corresponds to the limit of validity of the linearized equations. Yet, in order to compute $\avg{|G_{ii}|^p}$ we need to take into account the region of large  values of the imaginary part of the Green's functions and not the region of typical finite values. It seems therefore that the linearized equations are not useful. Luckily enough, this is not the case. To see this we define:
\begin{equation} \label{eq:mii1}
{\cal M}_{ii} \equiv \frac{1}{G_{ii}} = \epsilon_{i} - {\rm i} \, \eta - \!\! \sum_{m \in \partial i} G_{m \to i} = m_{ii} - {\rm i} \eta \, \hat{m}_{ii} \, , 
\end{equation}
from which  one immediately obtains that 
\[
\langle |G_{ii}|^p \rangle  = \int  Q(m,\hat{m}) \frac{1}{(m^2+\hat{m}^2 \eta^2)^{p/2}}   \, \de m \, \de \hat{m} \, .
\]
Similarly, using the fact that $\hat{g}_{ii} = \hat{m}_{ii}/(m_{ii}^2 + \eta^2 \hat{m}_{ii}^2)$, $\langle {\rm Im} G_{ii} \rangle$ is expressed as:
\[
\langle {\rm Im} G_{ii} \rangle = \int  Q(m,\hat{m}) \frac{\eta {\hat{m}}}{m^2+\hat{m}^2 \eta^2}   \, \de m\, \de \hat{m} \, ,
\]
Given that $\hat{m}$ is strictly positive we can make the change of variables $m = \eta \hat{m} x$ that leads to
\begin{eqnarray} 
\langle |G_{ii}|^p \rangle &=& \int  Q(\eta \hat{m} x  ,\hat{m}) 
\frac{(\eta \, \hat{m})^{1-p}}{(1 + x^2)^{p/2}}   \, \de x \, \de \hat{m} \, , \label{eq:gii} \\
\label{eq:img} \langle {\rm Im} G_{ii} \rangle &=&  \int  Q(\eta \hat{m} x  ,\hat{m}) \frac{1}{1 + x^2}   \, \de x \, \de \hat{m} \, .
\end{eqnarray}
In the $\eta \to 0$ limit we can approximate $Q(\eta \hat{m}  x,\hat{m}) \approx Q(0,\hat{m})$ and perform the integration over $x$ explicitly. From Eq.~\eqref{eq:img} we immediately obtain that the average DoS~\eqref{eq:DoS} is given by:
\begin{equation} \label{eq:rhoav}
    \avg{\rho} = \int  Q(0 ,\hat{m})  \, \de \hat{m} \, .
\end{equation}
Plugging Eqs.~\eqref{eq:gii} and \eqref{eq:img} into Eq.~\eqref{eq:Ip}, one finally obtains:
\begin{equation} \label{eq:IpQ}
    I_p = 
    \avg{\rho}^{-1} \int  Q(0  ,\hat{m}) \, \hat{m}^{1-p} \,\de \hat{m} \, .
\end{equation}
A similar expression has been derived in Refs.~\cite{mirlin1994distribution,mirlin1994statistical} in the supermatrix NL$\sigma$M framework. We will discuss this connection in Sec.~\ref{sec:Tu}. 

To sum up, although the moments of the local Green's functions are controlled by the fact that  $|G_{ii}|$ is $O(1/\eta)$ with probability $O(\eta)$, they can be computed in terms of
the {\it typical} values of ${\cal M}_{ii}$, whose real part that is typically $O(1)$ and whose imaginary part that is typically $O(\eta)$.
The fact that in the localized phase one can use the linearized equations to compute the relevant observables, such as the (generalized) IPR, facilitates the adoption of highly efficient computational methods that strongly reduces the effect of the finite size of the population compared to the delocalized phase.

\begin{figure*}
\includegraphics[width=0.42\textwidth]{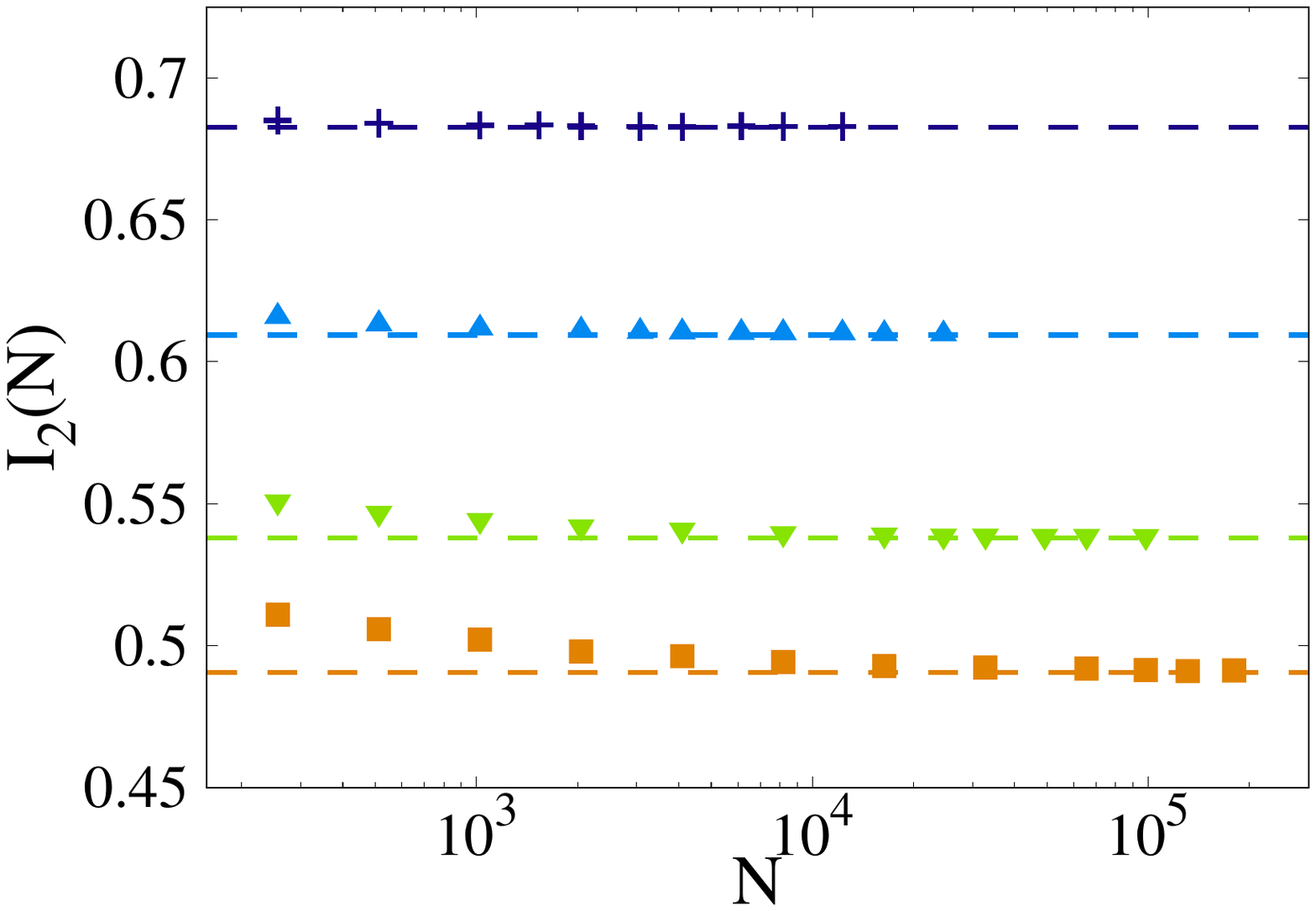} \hspace{0.5cm} 
\includegraphics[width=0.42\textwidth]{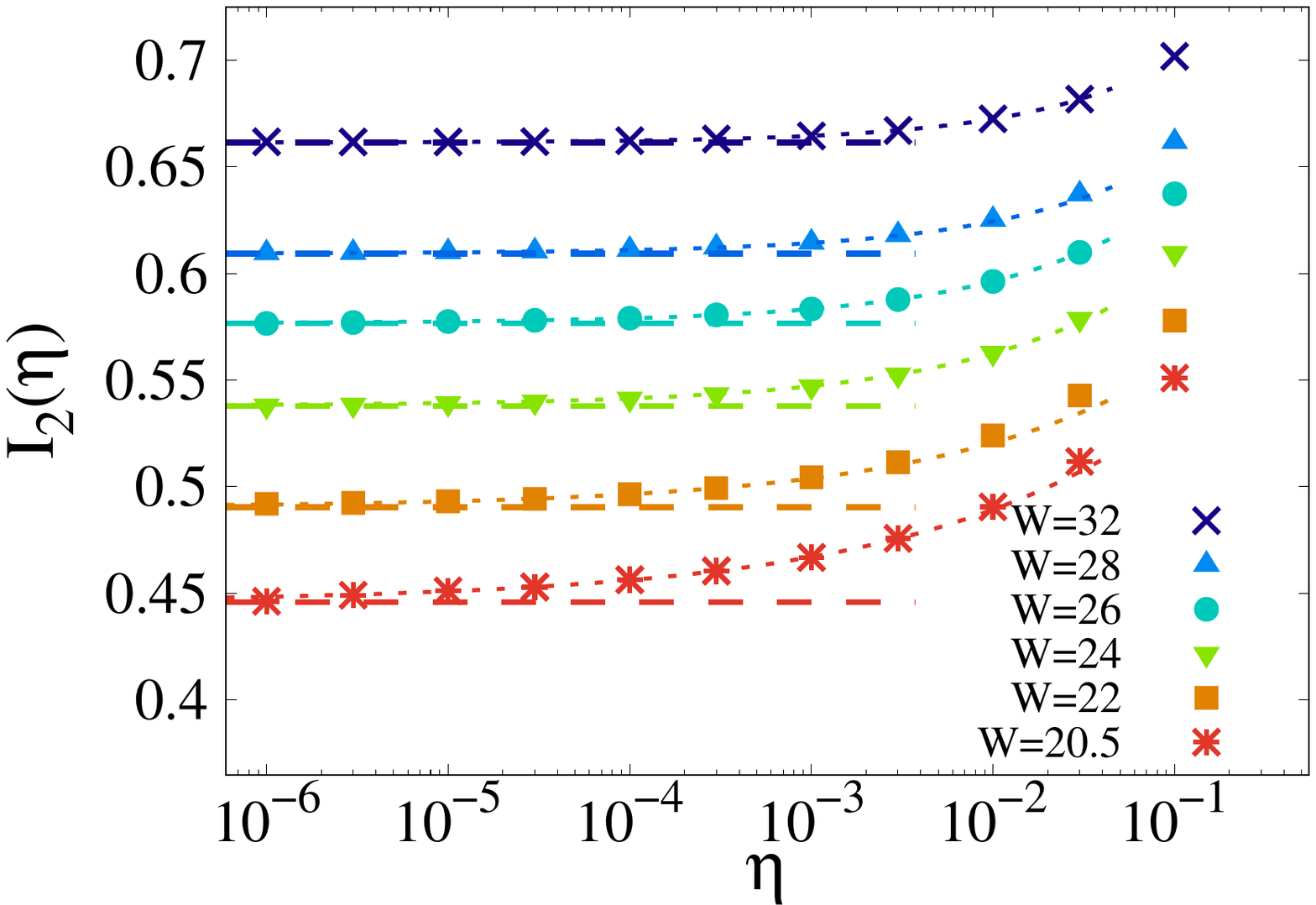} 
\caption{Left: Average IPR of the eigenvectors of the Anderson model on a RRG  of $N$ nodes close to zero energy (see App.~\ref{app:ED}) for $W=34$ ($+$), $W=28$ ($\blacktriangle$), $W=24$ ($\blacktriangledown$), and $W=22$ ($\blacksquare$). The horizontal dashed lines correspond to the asymptotic values of $I_2$ on the infinite BL found from Eq.~\eqref{eq:IpQ} using the computational method described in Sec.~\ref{sec:algo}. (The symbols' size exceeds that of the error bars.) Right: IPR as a function of the imaginary regulator $\eta$, $I(\eta) \equiv \eta \avg {|G_{ii}|^2}/\avg{\Im G}$,  computed solving the non-linearized cavity equations~\eqref{eq:cavity} with the standard population dynamics algorithm at finite $\eta$ and for several values of $W$. The dotted curves are fits of the data of the form $I_2 (\eta) \simeq I_2(\eta=0) + a_\eta \eta^{b_\eta}$. The horizontal dashed lines show the value of the IPR computed directly in the $\eta \to 0$ limit from Eq.~\eqref{eq:IpQ}, $I_2(\eta=0)$, using the procedure described in the text. The exponent $b_\eta$ goes to zero at $W_c$ as $b_\eta \propto (W-W_c)^{\kappa}$ with $\kappa \simeq 0.4 \pm 0.06$, while the coefficient $a_\eta$ is of $O(1)$ and does not vary significantly with $W$. (The symbols' size exceeds that of the error bars.)
\label{fig:IPR}}
\end{figure*}

\subsection{An efficient computational scheme for $Q(0,\hat{m})$} \label{sec:algo}

Here we introduce a modification of the population dynamics algorithm which allows us performing the extrapolation of $Q(m,\hat{m})$ to $m=0$ very efficiently, thereby allowing one to evaluate Eqs.~\eqref{eq:rhoav} and~\eqref{eq:IpQ} with arbitrary accuracy.  In fact, from Eq.~\eqref{eq:mii1} we have that $m_{ii} = \epsilon_{i} - \sum_{m \in \partial i} g_{m \to i}$. Hence, the probability that $m_{ii}=0$ is equal to the probability that $\epsilon_i = \sum_{m \in \partial i} g_{m \to i}$. This occurs with probability density $1/W$  if $|\sum_{m \in \partial i} g_{m \to i}|<W/2$, and with zero probability otherwise. Based on this observation, we thus proceed in the following way:
For a given value of $W > W_c$, we implement the standard population dynamics algorithm described in Sec.~\ref{sec:model} and obtain the stationary probability distribution of the cavity Green's function $P(g,\hat{g})$ in the linearized regime, corresponding to the solution of Eqs.~\eqref{eq:real} and~\eqref{eq:imag}; We extract $k+1$ elements $(g_\alpha,\hat{g}_\alpha)$ from the population and compute $m$ and $\hat{m}$ from Eq.~\eqref{eq:mii1}. We define $S = \sum_{\alpha=1}^{k+1} g_{\alpha}$; If (and only if) $|S|<W/2$ we add $\hat{m}^{1-p}/W$ to the numerator and $1/W$ to the denominator of $I_p$; We repeat this process several times and divide the numerator and the denominator by the total number of attempts; We renew the elements of the pool of the cavity Green's function by performing a few steps of the standard population dynamics algorithm and repeat the whole process several times until the desired accuracy on $I_p$ is reached. It is worth to mention that the algorithm described here, which is schematically summarized in App.~\ref{sec:algoIp}, can be straightforwardly extended to the computation of generic two-points correlation functions.

\subsection{Numerical results for the generalized IPRs} \label{sec:IPRsquareroot}

Below we present the numerical results obtained applying the procedure described above. (All the results presented in this paper are obtained with pools of $\Omega = 2^{28}$ elements.) We start by focusing on the full probability distribution of $\hat{m}$ when $m$ is identically equal to zero (divided by $\avg{\rho}$ to normalize it to $1$): 
$Q(0,\hat{m})/\avg{\rho}$. These probability distributions are plotted in the left panel of Fig.~\ref{fig:Pmhat} for several values of the disorder close to the critical point, $W_c \approx 18.17$~\cite{tikhonov2019critical}. The figure shows the appearance of the power-law tails at large $\hat{m}$ with a disorder-dependent exponent $\beta$, as expressed in Eq.~\eqref{eq:ansatz}. The values of $\beta$ extracted from the fit of the tails is reported in the right panel of Fig.~\ref{fig:Pmhat}. The dashed line represents the prediction of Eq.~\eqref{eq:beta} obtained from the direct diagonalization of the integral operator~\eqref{eq:Kernel} close to the critical point performed in~\cite{tikhonov2019critical}, which is in excellent agreement with the numerical results. 

 In the left panel of Fig.~\ref{fig:IPR} we explicitly check that the IPR measured from exact diagonalizations of RRGs of $N$ nodes (see App.~\ref{app:ED} for more details) converges in the large $N$ limit 
 to the values obtained using the ``improved'' population dynamics scheme described in Sec.~\ref{sec:algo}. 
 Yet, upon decreasing $W$ towards the critical point, the finite-size corrections to the asymptotic value becomes stronger and one needs to diagonalize larger systems in order to see the convergence. 
 As a consequence, a precise estimation of the IPR sufficiently close to $W_c$ from exact diagonalizations of finite-size samples is practically out of reach. Concretely, with currently available resources one cannot get reliable results for $W \lesssim 22$. (The finite $N$ corrections of the IPR to the $N \to \infty$ value will be studied in detail in a forthcoming work.)
 
 In the right panel of Fig.~\ref{fig:IPR} we show that the IPR obtained using standard population dynamics for the (non-linearized) self-consistent cavity equations~\eqref{eq:cavity} in presence of a small but finite imaginary regulator converges in the small $\eta$ limit to the value obtained directly at $\eta=0$ from Eq.~\eqref{eq:IpQ} using the algorithm described in Sec.~\ref{sec:algo}. However, as $W$ gets closer to $W_c$ the finite-$\eta$ corrections become stronger and one needs to consider smaller and smaller $\eta$ to see convergence: The data are well fitted by $I_2 (\eta) \simeq I_2(\eta=0) + a_\eta \eta^{b_\eta}$ (dashed lines), with an exponent $b_\eta$ decreasing with $W$ and  approaching zero at $W_c$ as $b_\eta \propto (W-W_c)^{\kappa}$. 
 Since upon decreasing $\eta$ the probability distribution of the imaginary part of the Green's functions becomes broader and broader, obtaining accurate estimations for its moments, which are controlled by the tails, becomes increasingly hard. In practice, using the standard population dynamics algorithm  one can measure $\eta \avg{|G_{ii}|^2}$ precisely enough  only for $\eta \gtrsim 10^{-7}$. For these reasons, computing the $\eta \to 0$ limit of the IPR close enough to $W_c$ with the standard approach is essentially unfeasible.  

\begin{figure*}
\includegraphics[width=0.42\textwidth]{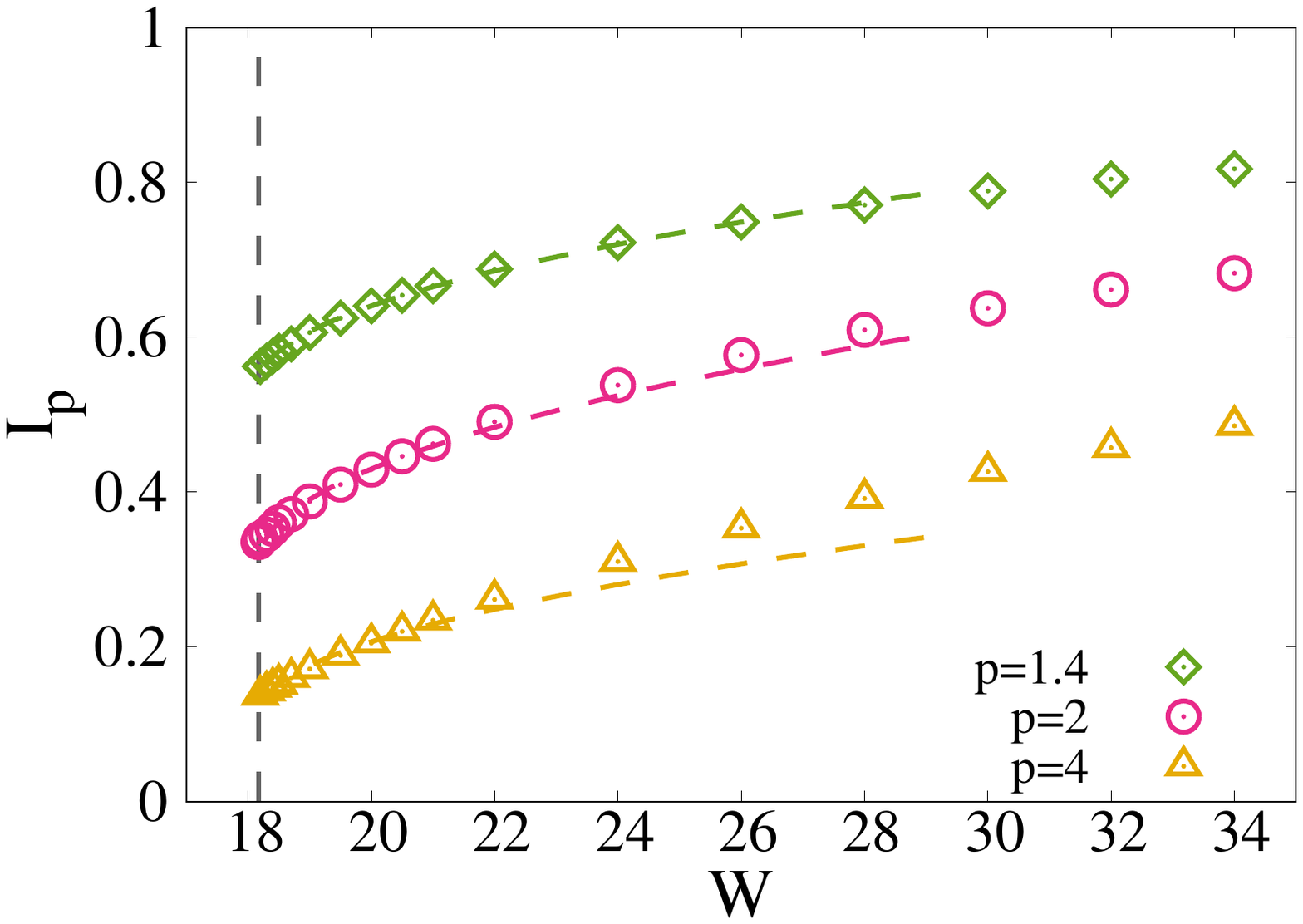} \hspace{0.5cm} \includegraphics[width=0.42\textwidth]{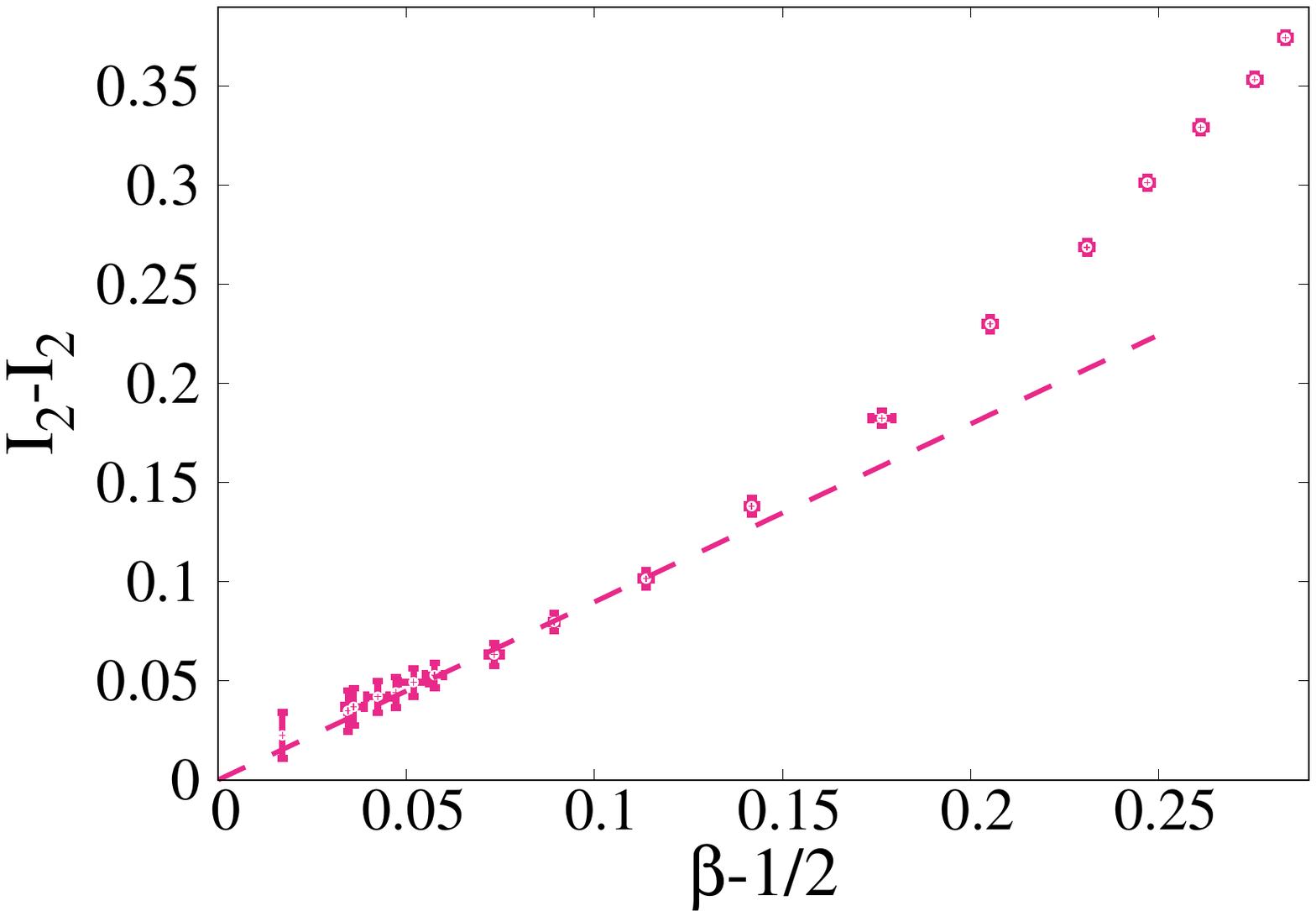} 
\put(-215,88){\rotatebox[origin=c]{90}{\small $(c)$}}  
\caption{Left: Generalized inverse participation ratios $I_p$ (for $\eta=0$ and $N \to \infty$) as a function of $W$ in the localized phase for $p=1.4$, $p=2$, and $p=4$. The dashed lines are fits of the data close to $W_c$ of the form $I_p \simeq I_p^{\rm (c)} + a_p \sqrt{W - W_c}$. We find $I_{1.4}^{\rm (c)} \simeq 0.539$ and $a_{1.4} \simeq 0.075$, $I_2^{\rm (c)} \simeq 0.304$ and $a_2 \simeq 0.094$, and $I_4^{\rm (c)} \simeq 0.112$ and $a_4 \simeq 0.07$. (The symbols' size exceeds that of the error bars.) Right: Parametric plot of $I_2 - I_2^{(c)}$ as a function of $\beta-1/2$ (with $I_2^{(c)} = 0.304$). The dashed linear is a fit of the data close to the critical point of the form: $I_2 - I_2^{(c)} = \alpha (\beta - 1/2)$ with $\alpha \approx 0.882$.
\label{fig:IPRc}}
\end{figure*}

Finally, we specifically focus on the critical behavior of the (generalized) IPR close to $W_c$. In Fig.~\ref{fig:IPRc}(left) we plot $I_p$ (for $\eta=0$ and $N\to \infty$) as a function of $W\ge W_c$, for $p=1.4$, $p=2$, and $p=4$, showing that $I_p$ jumps to a finite value at $W_c$, as predicted by the analytic solution~\cite{efetov1985anderson,efetov1987density,efetov1987anderson,zirnbauer1986localization,zirnbauer1986anderson,verbaarschot1988graded,mirlin1991localization,mirlin1991universality,fyodorov1991localization,fyodorov1992novel,mirlin1994statistical,tikhonov2019statistics}. The behavior of $I_p$ for $W \gtrsim W_c$ is well described by:
\begin{equation} \label{eq:IPRsingularity}
    I_p \simeq I_p^{\rm (c)} + a_p \sqrt{W - W_c} \, . 
\end{equation}
Specifically, for $p=2$ we find $I_2^{\rm (c)} \simeq 0.304$ and $a_2 \simeq 0.094$. 
To support this claim, in the right panel of Fig.~\ref{fig:IPRc} we perform a parametric plot of $I_2 - I_2^{(c)}$ as a function of $\beta-1/2$, showing that close enough to the localization transition the data are well described by a linear relation. 

The confirmation of the jump in the IPR's at the localization transition, as predicted by the supersymmetric analysis~\cite{efetov1985anderson,efetov1987density,efetov1987anderson,zirnbauer1986localization,zirnbauer1986anderson,verbaarschot1988graded,mirlin1991localization,mirlin1991universality,fyodorov1991localization,fyodorov1992novel,mirlin1994statistical,tikhonov2019statistics}, is an important result, particularly because it has been challenged in recent studies~\cite{garcia2022critical} (see also~\cite{chen2023quantum}). Moreover, the square root singularity identified in Fig.~\ref{fig:IPRc}(left)---standing out as one of the most significant contributions of our work---has not been previously documented in the literature.
Exploring the possible connection between this distinctive behavior and the recently emphasized transverse length's singular behavior~\cite{garcia2020two,garcia2022critical}---which governs the exponential decay of wave-functions along typical branches of the tree---would offer intriguing insights on the geometric structure of Anderson localized eigenstates on the BL. 

To conclude this section, it is worth mentioning that by extending the analysis to higher values of the connectivity of the BL (not shown), we observe that the amplitude of the jump of the IPR at $W_c$ grows with $k$ and appears to approach $1$ in the infinite connectivity limit, as predicted in~\cite{mirlin1997localization}.

\begin{figure*}
\includegraphics[width=0.33\textwidth]{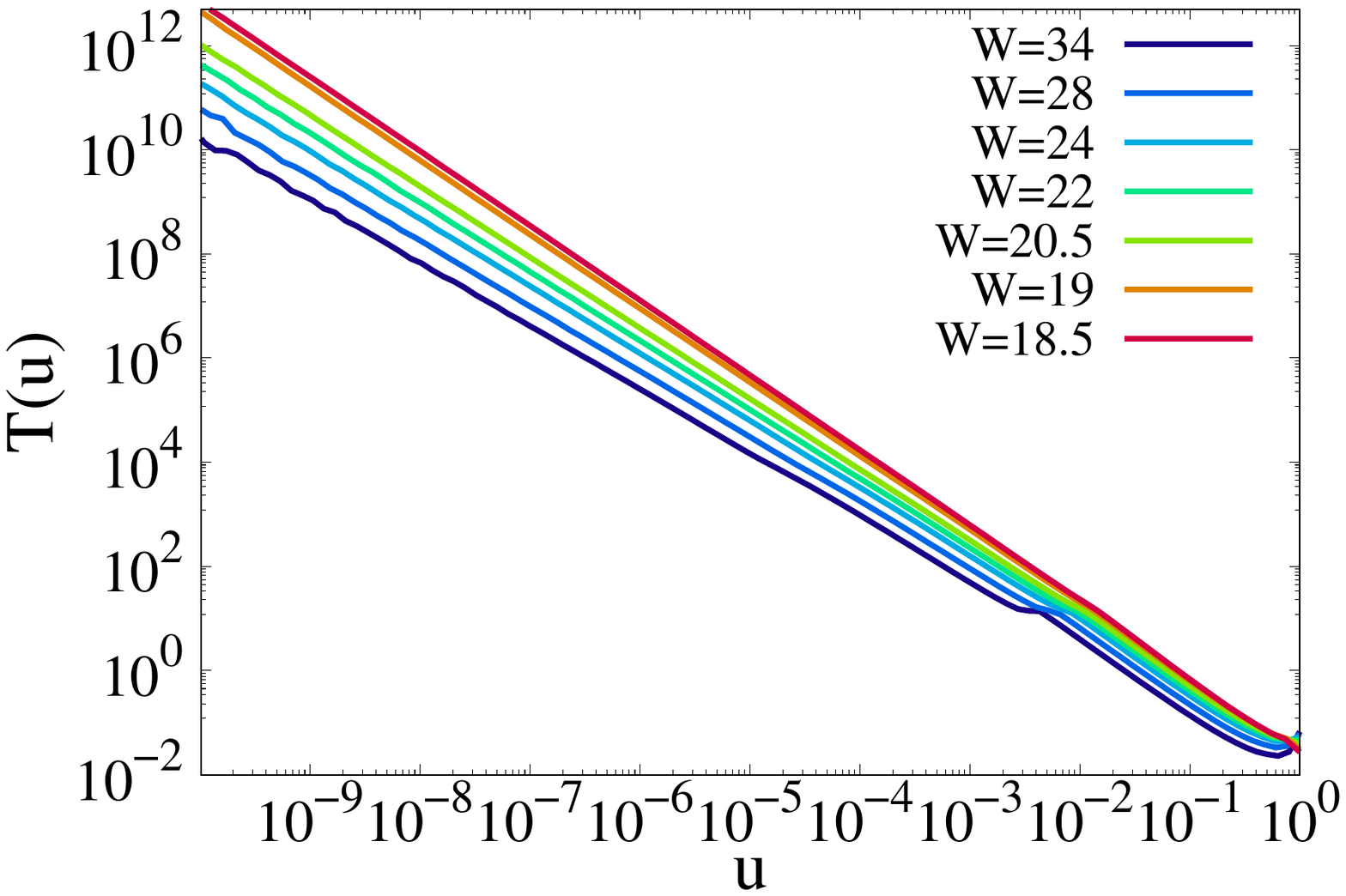} \hspace{-0.2cm} \includegraphics[width=0.33\textwidth]{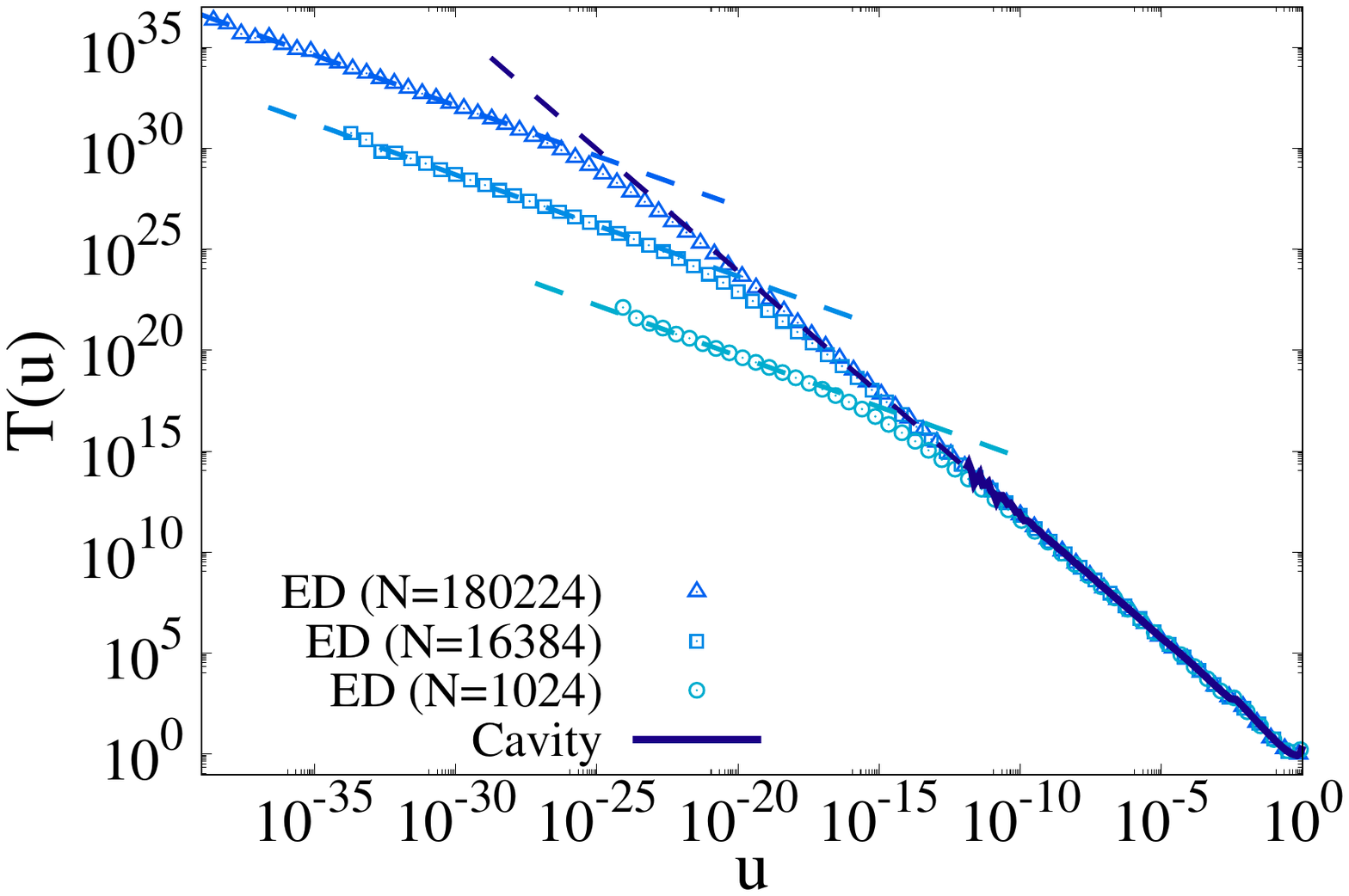} \hspace{-0.2cm} \includegraphics[width=0.33\textwidth]{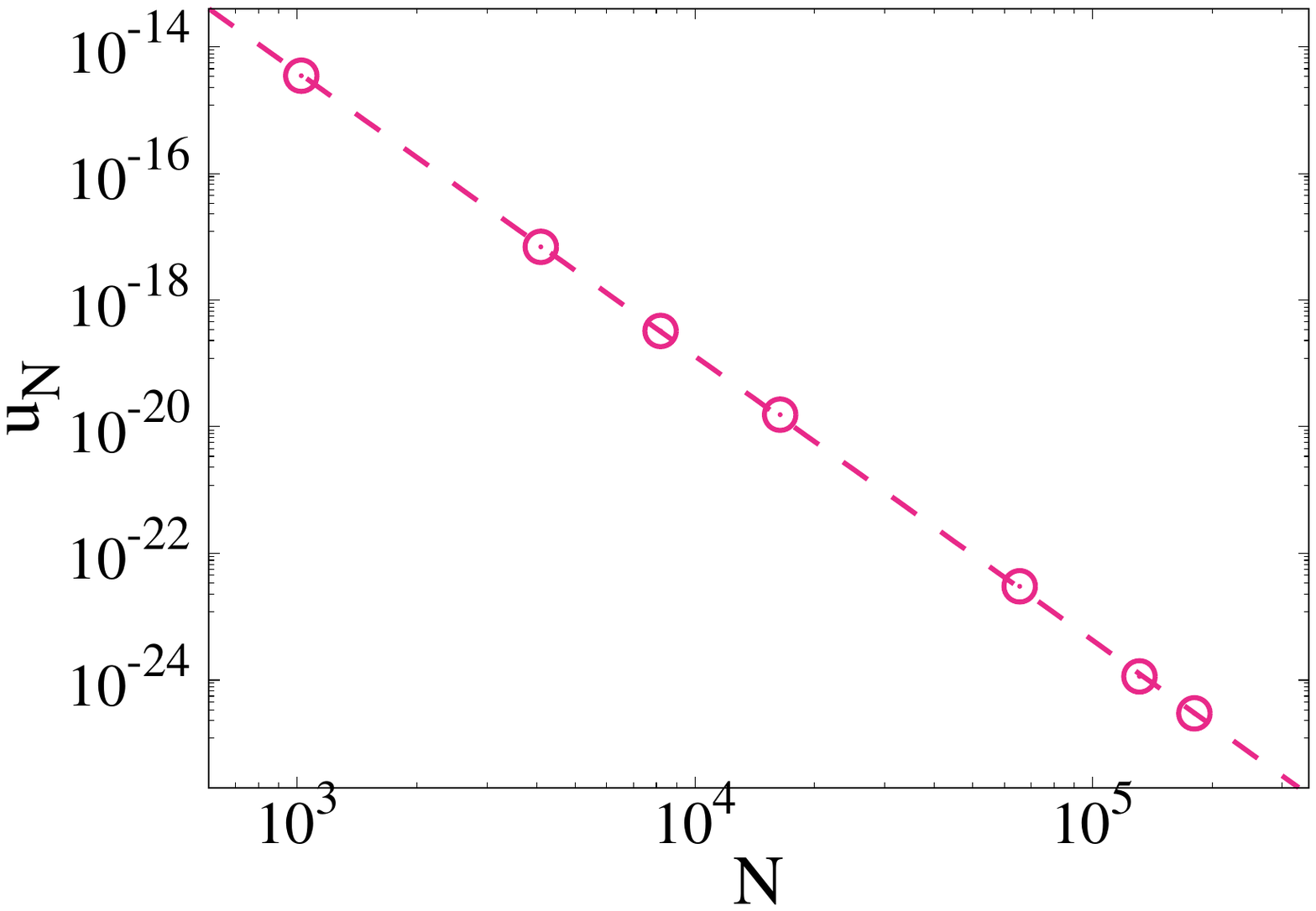}
\caption{Left: Probability distributions of the wave-functions' amplitudes $T(u)$, Eqs.~\eqref{eq:Tu} and~\eqref{eq:Tu1}, for several values of $W$ across the localized phase computed within the ``improved'' population dynamics scheme described in Sec~\ref{sec:algo}. Note that $T(u)$ is normalized in such a way that $\int \! T(u) \, \de u = N$. For small $u$ the distributions behave as $T(u) \propto u^{2 - \beta}$, Eq.~\eqref{eq:TuPL}, with the exponent $\beta$ plotted in the right panel of Fig.~\ref{fig:Pmhat}. Middle: Comparison between the  distributions of the wave-functions' amplitudes obtained from the ``improved'' population dynamics 
(continuous curve), 
with the results of exact diagonalizations on RRGs of $N$ nodes (symbols) for $W=34$. For $u>u_N$ the distributions of the amplitudes measured on finite systems coincide with the one obtained within the cavity approach on the infinite BL, featuring a power-law behavior described by Eq.~\eqref{eq:TuPL} with $\beta \approx 0.784$. For $u<u_N$, instead, the distributions for RRGs of finite size $N$ crossover to an integrable square root singularity (dashed lines). Right: Crossover scale, $u_N$, as a function of the system size. The dashed line shows the power-law dependence of the cut-off as $u_N \propto N^{-1/(1-\beta)}$, in perfect agreement with the prediction of Eq.~\eqref{eq:uN}. (The symbols' size exceeds that of the error bars.)}
\label{fig:distributions}
\end{figure*}

\subsection{The Distribution function of the eigenfunctions' amplitudes} \label{sec:Tu}

Comparison of Eq.~\eqref{eq:giip} with Eq.~(\ref{eq:gii}) and averaging immediately leads to:
\begin{equation} \label{eq:moments}
   \left \langle \sum_{\alpha} |\psi_\alpha(i)|^{2 p} \delta(E_\alpha) \right \rangle = \int  Q(0,\hat{m}) \, \hat{m}^{1-p}  \, \de \hat{m} \, .
\end{equation}
This implies that the moments of the wave-functions' amplitudes can be directly expressed in terms of the distribution of the imaginary part of $G_{ii}^{-1}$ on the scale $\eta$. More precisely introducing the distribution $T(u)$ as in~\cite{mirlin1994statistical}:
\begin{equation} \label{eq:Tu}
T(u) \equiv   
   \frac{1}{\avg{\rho}} \left \langle  \sum_{\alpha} \delta \left ( u-|\psi_\alpha (i) |^{2} \right ) \delta(E_\alpha)\right \rangle \, ,
\end{equation}
Eq.~\eqref{eq:moments} leads to:
\begin{equation} 
T(u) = 
\avg{\rho}^{-1} Q\left(0,\frac{1}{u}\right) \frac{1} {u^3}  \, .
\label{eq:Tu1}
\end{equation}
In Refs.~\cite{mirlin1994distribution,mirlin1994statistical} the authors obtained the following expressions for the $T(u)$ and $I_p$ analogous to Eqs.~\eqref{eq:IpQ} and~\eqref{eq:Tu1}.
\begin{equation}
\begin{aligned}
    T(u) & = \frac{\de^2 F_l (u)}{\de u^2} \, , \\
    I_p & = p (p-1) \int u^{p-2} F_l(u) \, \de u \, ,
    \end{aligned}
\end{equation}
in terms of a function $F_l(u)$. By comparison one easily sees that $F_l(u)$ is related to our $Q(0,\hat{m})$ through
\begin{equation}
Q\left(0,\frac{1}{u}\right) = \avg{\rho} u^3  \frac{\de^2 F_l (u)}{\de u^2} \, .
\end{equation}
The numerical results for the distributions $T(u)$ are shown in the left panel of Fig.~\ref{fig:distributions} for several values of the disorder across the localized phase. 
Since $Q(0,\hat{m})$ goes to zero as $1/\hat{m}^{1+\beta}$ for large $\hat{m}$ we have
\begin{equation} \label{eq:TuPL}
T(u) \propto \frac{1}{u^{2-\beta}} \, .
\end{equation}
From the above expression one has that $\int \! T(u) \, \de u$ is divergent for small values of $u$. On the other hand this is not consistent with the fact that $\int T(u) \de u = N$ exactly by definition. Ref.~\cite{mirlin1994statistical} argues that the matching between the finite $N$ result and thermodynamic limit expression (\ref{eq:Tu1}) occurs because the integral must be truncated at a some value $u_N$ such that $\int_{u_N}^{\infty} T(u) \, \de u = N$ and this leads naturally to 
\begin{equation} \label{eq:uN}
u_N \propto \frac{1}{N^{\frac{1}{1-\beta}}} \, .   
\end{equation}
This phenomenon is clearly illustrated in the middle and right panels of Fig.~\ref{fig:distributions} for $W=34$ (similar results, not shown, are found for other values of $W$ within the localized phase). In the middle panel we plot the probability distributions of the wave-functions' amplitudes computed  from exact diagonalizations of finite RRGs of $N$ nodes (see App.~\ref{app:ED}). For $u>u_N$ these distributions coincide with the one obtained using the cavity approach on the infinite BL, and feature a power-law behavior given in Eq.~\eqref{eq:TuPL} with $\beta \approx 0.784$. For $u \simeq u_N$ the probability distributions on finite graphs exhibit a crossover to a different behavior  
at small amplitudes described by an integrable square root singularity, $T(u) \propto 1/\sqrt{u}$. The position of the crossover moves to smaller values when the system size is increased, in agreement with the arguments given above. This is shown in the right panel, which indicates that the dependence of the crossover $u_N$ upon the system size is very well described by Eq.~\eqref{eq:uN}.

\begin{figure*}
\includegraphics[width=0.333\textwidth]{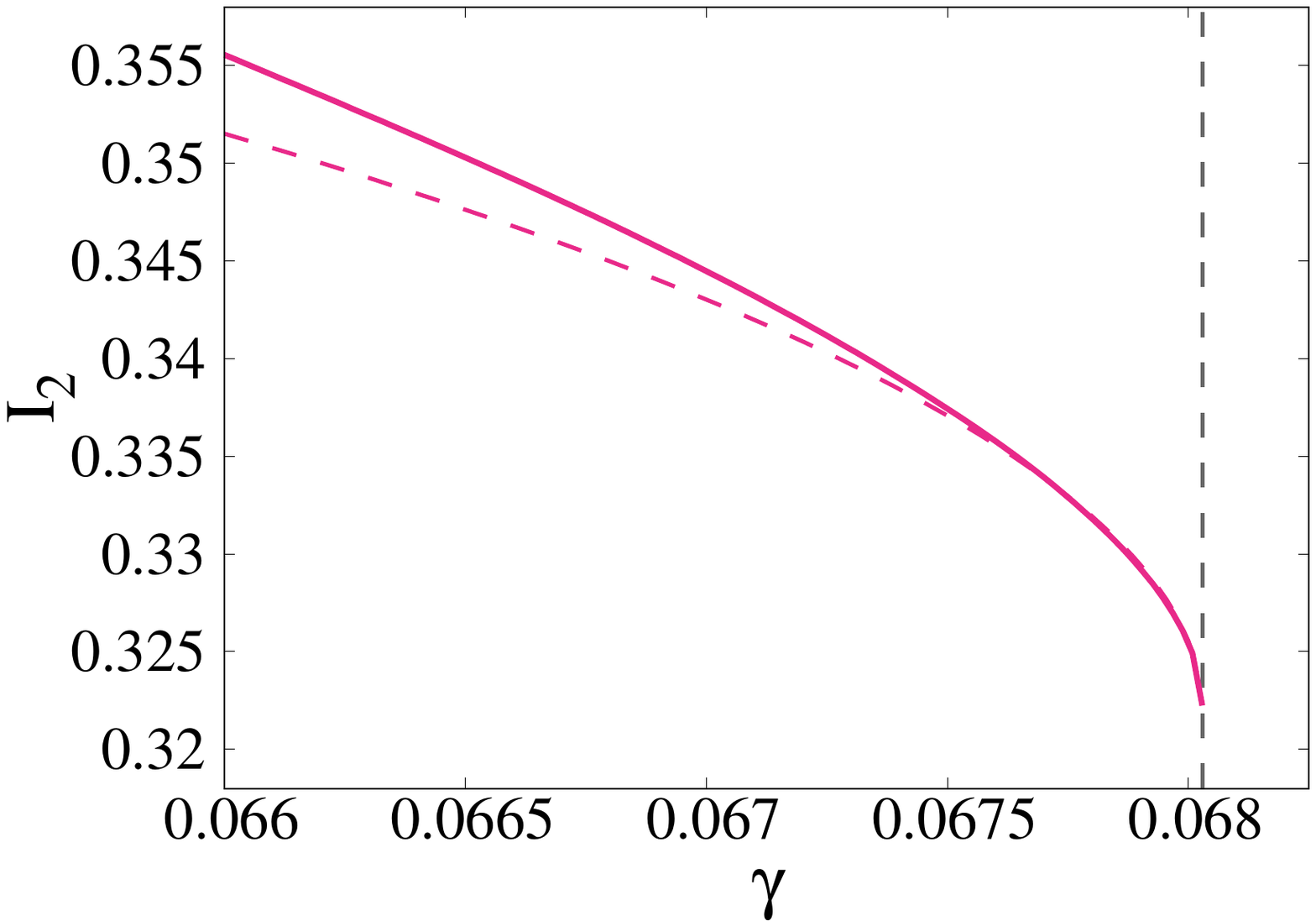} \hspace{-0.2cm} \includegraphics[width=0.333\textwidth]{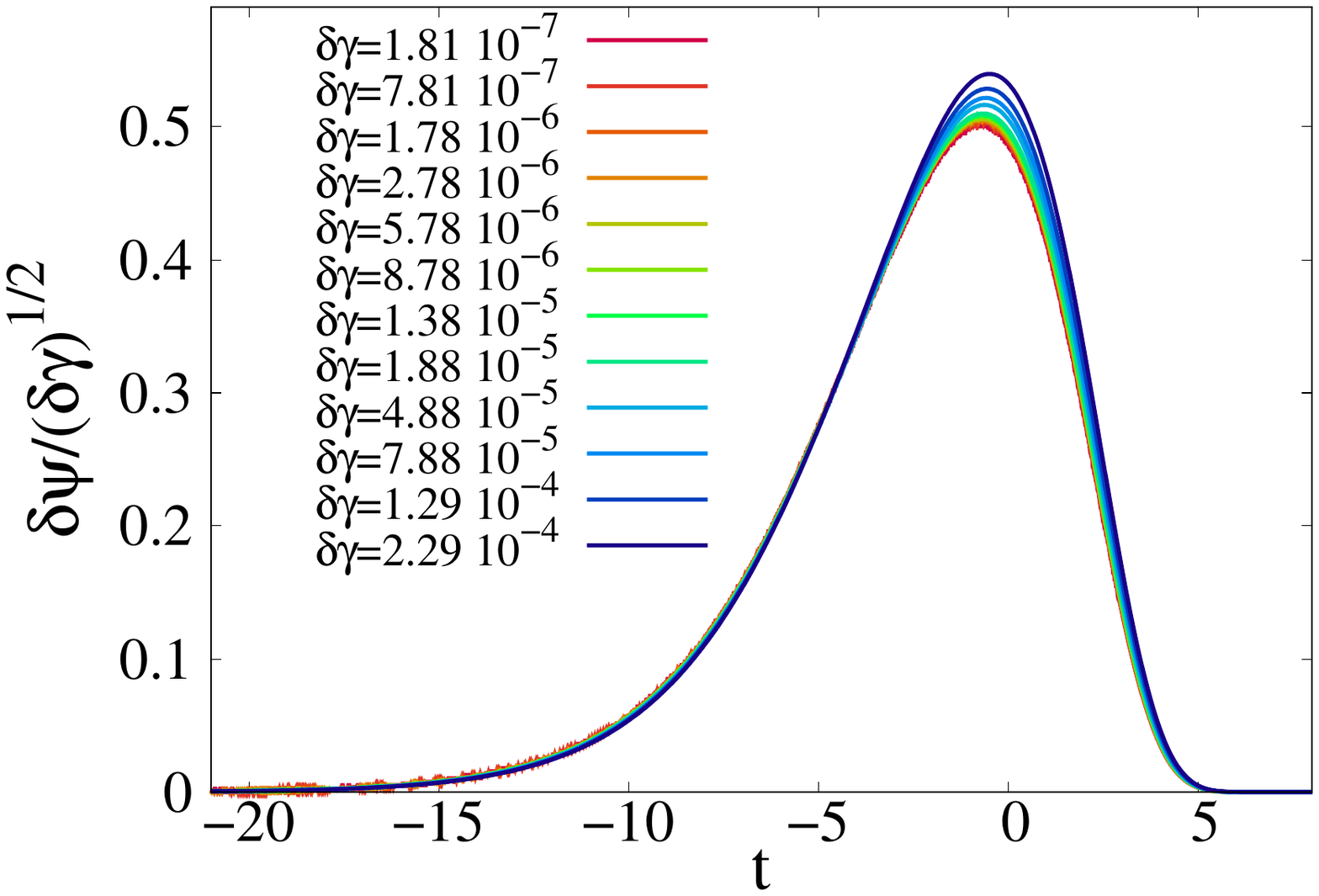} \hspace{-0.2cm} \includegraphics[width=0.333\textwidth]{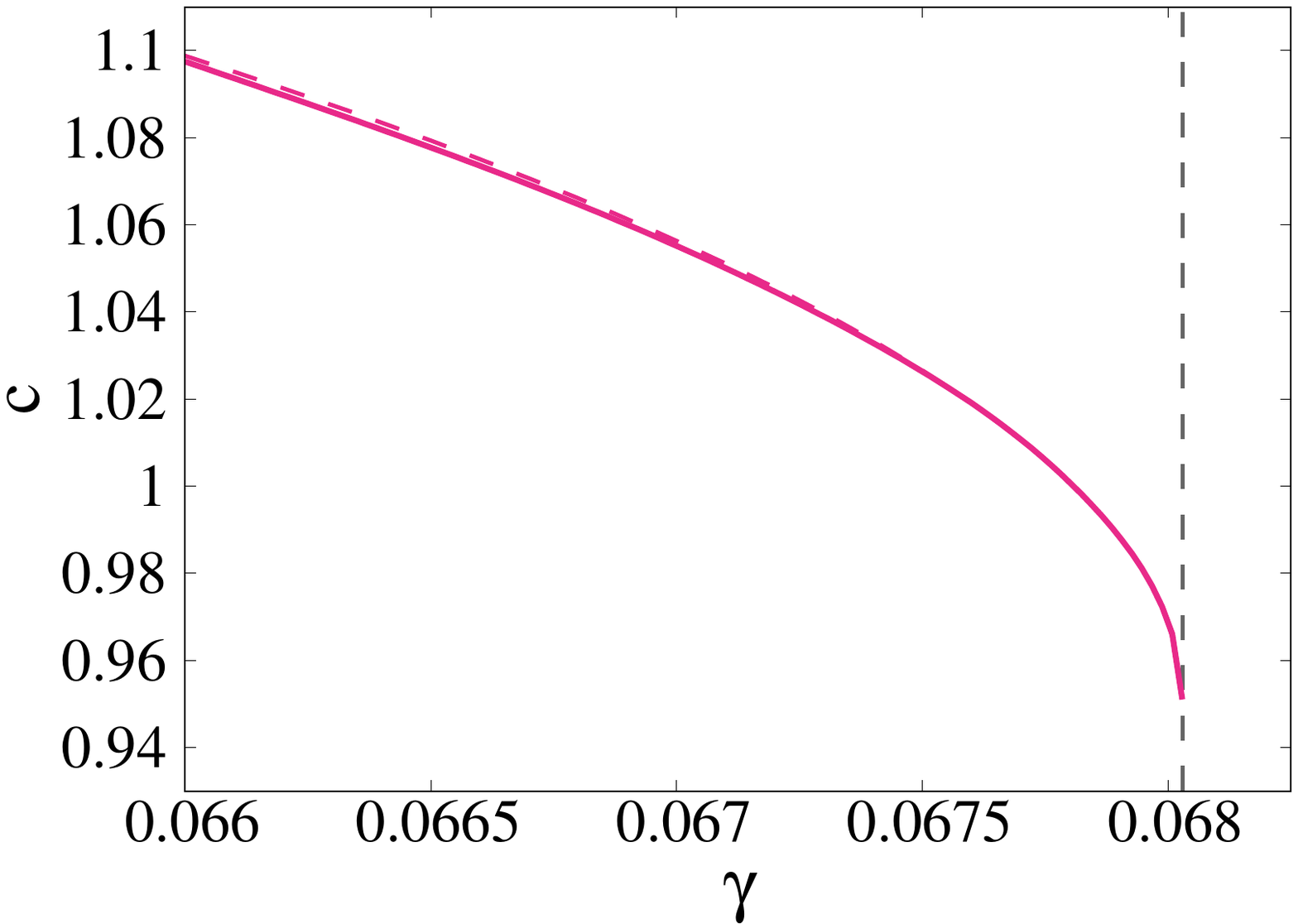} 
\caption{Left: Inverse participation ratio $I_2$, Eq.~\eqref{eq:I2nlsm}, as a function of $\gamma$ in the localized phase ($\gamma \le \gamma_c$) obtained within the NL$\sigma$M effective description of AL for $k=2$. The critical point is located at $\gamma_c \simeq 0.06803$ (vertical dashed line). The dashed curve is a fit of the data close to $\gamma_c$ of the form $I_2  = I_2^{(c)} + a \sqrt{\gamma_c - \gamma}$, with $I_2^{(c)} \approx 0.322$ and $a \approx 0.655$. Middle: Scaling plot of $\delta \psi (t;\gamma) \equiv \psi(t; \gamma) - \psi_c (t)$ divided by the square root of the distance from the transition point, $\sqrt{\delta \gamma} \equiv \sqrt{\gamma_c - \gamma}$, as a function of $t$ (where $\psi(t; \gamma)$ is the solution of Eqs.~\eqref{eq:NLsM}). Right: Prefactor $c$ of the exponential tail of $\psi(t)$, Eq.~\eqref{eq:psi_tail}, as a function of $\gamma$. The dashed line is a fit of the data of the form $c(\gamma) \simeq c_c + c_0 \sqrt{\gamma_c - \gamma}$ (with $c_c \approx 0.952$ and $c_0 \approx 3.28$). The horizontal dashed line signals the position of the transition point.
\label{fig:NLsM}}
\end{figure*}

\subsection{Singular behavior of the IPR within the NL$\sigma$M formulation}
In order to confirm the singular behavior of the IPR described in Sec.~\ref{sec:IPRsquareroot} and reported in Fig.~\ref{fig:IPRc} and to understand its origin, in this section we consider an effective field-theoretical description of the localization transition first introduced in Ref.~\cite{schafer1980disordered}, in which AL was mapped onto a non-linear $\sigma$ model with non-compact symmetry. The NL$\sigma$M representation is obtained as the $n \to \infty$ limit of an $n$-orbital generalization of the problem, which can be viewed as describing an electron hoping between metallic granules containing $n$ orbitals and located at the nodes of the same Bethe lattice.  For $n=1$ the Anderson tight-binding model (with random hopping $t_{ij}$) is recovered. Its $n$-orbital generalization is expected to exhibit the same gross features and the same critical behavior, with the advantage that analytical calculations are usually somewhat simpler. The NL$\sigma$M on an infinite BL was solved via the supersymmetry approach in Refs.~\cite{efetov1985anderson,efetov1987density,efetov1987anderson,zirnbauer1986localization,zirnbauer1986anderson,verbaarschot1988graded}. Such solution is expressed in terms of the following self-consistent integral equation for an order parameter function $\psi(t)$, which is essentially akin to the Laplace transform of the probability distribution of the imaginary part of the Green's functions (with the change of variable $t= \ln s$, $s$ being the variable of the Laplace transform):
\begin{equation} \label{eq:NLsM}
\begin{aligned}
    \psi(t) & = \int_{- \infty}^{+\infty} \de t^\prime \, L_\gamma (t - t^\prime) \, d(t^\prime) \, \psi^k (t^\prime) \, , \\
    d(t) &= \exp \left( -2 e^t \right ) \, , \qquad \,\,\,\,\,\,\,\,\, L_\gamma (t) = e^{t/2} \ell_\gamma (t) \, , \\
    \ell_\gamma(t) &= \left( \frac{\gamma}{2 \pi} \right)^{1/2} e^{- \gamma \cosh t} \bigg [ \sinh \gamma \cosh t +\\
    & \qquad \qquad \qquad \qquad \qquad \,\,\,\,\,\, + \left( \cosh \gamma - \frac{\sinh \gamma}{2 \gamma} \right)\bigg] \, .
    \end{aligned}
\end{equation}

In the NL$\sigma$M formulation the parameter $\gamma$ is a dimensionless coupling constant, which plays the role of $t_{ij}/W$. The solution of this equation vanishes at $t \to \infty$, due to the fact that $d(t \to \infty) = 0$, and goes to a constant for $t \to - \infty$  where $d(t \to - \infty) = 1$. In fact one can show~\cite{zirnbauer1986anderson,efetov1985anderson} that the solution decreases monotonically from $1$ to $0$ as $t$ varies from $-\infty$ to $+\infty$ and has a sharp kink in a region where it decreases rapidly. In the localized phase (\ie, small values of $\gamma$), the kink is located somewhere near $t=0$. For $\gamma$ larger than a critical value $\gamma_c$, instead, the kink is unstable and runs away to minus infinity. Thus the existence of a non-trivial solution of Eq.~\eqref{eq:NLsM} characterizes the localized phase, while a trivial solution ($\psi(t)=0$ for $t>-\infty$)  corresponds to the metallic phase. To find the limit of stability of the insulating phase one can consider  the linearized equation which describes the effect of infinitesimal perturbations on the solution $\psi(t) = 1$ for large negative $t$. This analysis leads to the study of the spectral properties of the kernel $L_\gamma$, whose largest eigenvalue $\lambda_\gamma (\beta)$ is given by~\cite{efetov1985anderson,zirnbauer1986anderson}:
\begin{equation} \label{eq:beta_nlsm}
    \lambda_\gamma (\beta) = \int_{- \infty}^{+\infty} \de t \, e^{(1/2 - \beta) t} \, \ell_\gamma (t) \, ,
\end{equation}
with $\beta \in [0,1]$. $\lambda_\gamma (\beta)$ shares the very same properties (discussed in Sec.~\ref{sec:critical}) of the largest eigenvalue of the integral operator defined by the Kernel~\eqref{eq:Kernel} which emerges in the Anderson problem when studying the stability of the linearized solution of the cavity equations in the localized phase with respect to a small imaginary part of the Green's functions. In particular $\lambda_\gamma (\beta)$ is symmetric for $\beta \to 1 - \beta$ and is thus minimal for $\beta=1/2$. Close to the transition $\beta$ behaves as $\beta \simeq 1/2 + {\rm cst} \sqrt{\gamma_c - \gamma}$. The fact that $\beta=1/2$ at the critical point then yields a closed equation for $\gamma_c$~\cite{zirnbauer1986anderson,efetov1985anderson}.  When the transition is approached from the localized side, $\gamma \lesssim \gamma_c$, the solution of Eqs.~\eqref{eq:NLsM} assumes the asymptotic form 
\begin{equation} \label{eq:psi_tail}
\psi(t) \simeq 1 - c \, e^{\beta t} \, , \qquad \qquad \textrm{(for $t \ll -1$)} \, ,
\end{equation}
where $c$ is a $\gamma$-dependent constant of order unity. The left exponential tail of $\psi(t)$ corresponds in fact to the power-law tails of $Q(0,\hat{m})$ at large $\hat{m}$ (see Eq.~\eqref{eq:ansatz} and Fig.~\ref{fig:Pmhat}(left)).

\begin{figure*}
\includegraphics[width=0.42\textwidth]{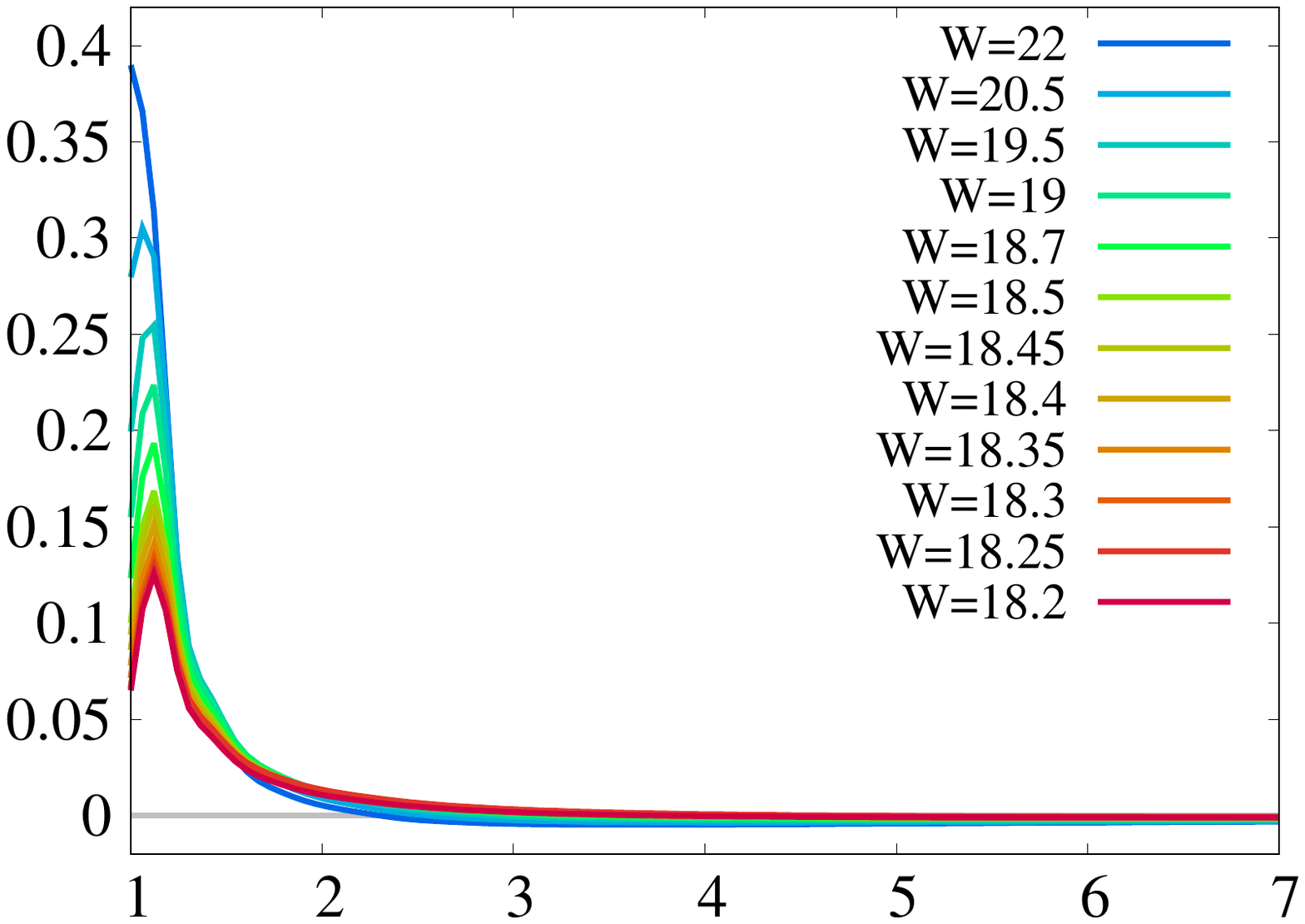} 
\put(-218,73){\rotatebox[origin=c]{90}{\large $\delta \tilde{Q}(\hat{m}) /(\delta W)^{1/2}$}} \put(-99,-8){\large $\hat{m}$}\hspace{0.5cm} \includegraphics[width=0.415\textwidth]{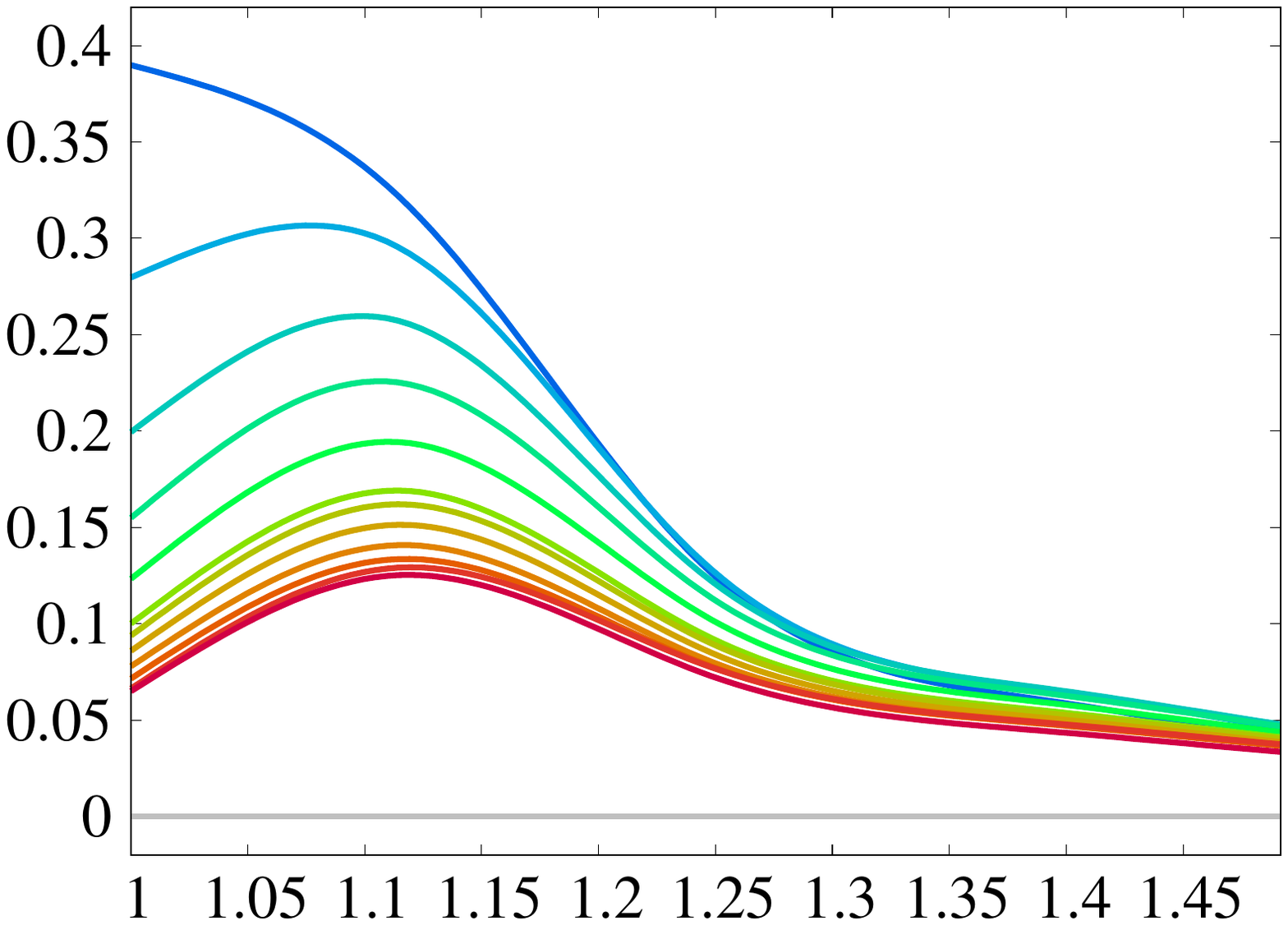} \put(-218,73){\rotatebox[origin=c]{90}{\large $\delta \tilde{Q}(\hat{m}) /(\delta W)^{1/2}$}} \put(-99,-9){\large $\hat{m}$}
\caption{ Scaling plot of $\delta \tilde{Q}(\hat{m};W)$, defined in Eq.~\eqref{eq:deltaQ}, divided by the square root of the distance from the transition point, $\sqrt{\delta W} \equiv \sqrt{W - W_c}$, as a function of $\hat{m}$. The right panel shows a zoom of the same curves in the region $\hat{m} \sim 1$.}
\label{fig:deltaQ}
\end{figure*}

We have solved Eqs.~\eqref{eq:NLsM} numerically for $k=2$ by iteration for several values of $\gamma \le \gamma_c$. For $k=2$ the critical point is located at $\gamma_c \simeq 0.06803$. In practice we discretized the integral over $\de t^\prime$ on a finite mesh of constant spacing of $N_{\rm bin}$ points in the interval $[t_{\rm min}, t_{\rm max}]$. The boundaries of the interval are chosen in such a way that $\psi(t) = 1$ for $t < t_{\rm min}$ and $\psi (t) = 0$ for $t> t_{\rm max}$ within the numerical accuracy. Furthermore, in the interval $t \in [t_{\rm min}, t_{\rm tail}]$ the function $\psi(t)$ is set to be equal to Eq.~\eqref{eq:psi_tail}, with $\beta$ obtained from the solution of Eq.~\eqref{eq:beta_nlsm}. The constant $c$ is fixed in a self-consistent way, by imposing the continuity of the logarithmic derivative of $\psi(t)$ for $t=t_{\rm tail}$. Below we show the results obtained for $N_{\rm bin} = 6144$, $t_{\rm min} = -64$, $t_{\rm tail} = -44$, and $t_{\rm max} = 16$. 

Given the solution $\psi(t)$ of Eq.~\eqref{eq:NLsM}, the IPR is obtained as~\cite{zirnbauer1986anderson}:
\begin{equation} \label{eq:I2nlsm}
    I_2 = 2 \int_{- \infty}^{+\infty} \de t \,  e^t \, \exp \left( -2 e^t \right ) \psi^{k+1} (t) \, .
\end{equation}
The numerical results for $I_2$ are reported in the left panel of Fig.~\ref{fig:NLsM}, showing that, as for the Anderson model, the IPR has a finite jump at the transition followed by a square root singularity, as in Eq.~\eqref{eq:IPRsingularity}. To understand the origin of such behavior, in the middle panel of Fig.~\ref{fig:NLsM} we plot the difference between the solution found at $\gamma \lesssim \gamma_c$ and solution found right at the critical point, $\psi_c (t)$: $\delta \psi (t;\gamma) \equiv \psi(t; \gamma) - \psi_c (t)$. Close to $\gamma_c$ one has:
\[
I_2(\gamma) - I_2^{(c)}  \simeq 2 (k+1) \int_{- \infty}^{+\infty} \!\!\!\!\!\! \de t \,  e^t \, \exp \left( -2 e^t \right ) \psi_c^k (t) \, \delta \psi (t;\gamma) \, .
\]
The function $\delta \psi (t;\gamma)$ is the largest in correspondence of the kink, which is located approximately around $t=0$ and whose position moves to the left as $\gamma$ is increased towards $\gamma_c$. Due to the term $e^t \exp \left( -2 e^t \right )$ in the equation above the IPR is also dominated by the region around the kink. Yet, as shown in the figure, the $\delta \psi (t;\gamma)$'s obtained at different $\gamma$ collapse on the same function when divided by $\sqrt{\gamma_c - \gamma}$, implying that $I_2(\gamma) - I_2^{(c)} \propto \sqrt{\gamma_c - \gamma}$ close to $\gamma_c$.

A further confirmation comes from the inspection of the $\gamma$ dependence of the prefactor $c$ of the left exponential tail of $\psi(t)$, Eq.~\eqref{eq:psi_tail}. As shown in the right panel of Fig.~\ref{fig:NLsM}, the prefactor behaves as $c(\gamma) \simeq c_c + c_0 \sqrt{\gamma_c - \gamma}$ (with $c_c$ and $c_0$ of $O(1)$), implying that 
for $t \ll -1$ one has: 
\begin{equation} \label{eq:deltapsitails}
\delta \psi (t;\gamma) \simeq \sqrt{\gamma_c - \gamma} \, e^{\frac{1}{2} t} \left( \delta_1 t + \delta_2 + O\left(t \sqrt{\gamma_c - \gamma} \right) \right) \, ,
\end{equation}
with $\delta_1$ and $\delta_2$ of order $1$.

The analysis of the NL$\sigma$M thus provides a clear mathematical explanation of the origin of the square root singularity of the IPR: Although the main contribution to $I_2$ comes from the region around the kink, 
the matching with the tails at $t \ll -1$ 
produces a scaling regime close to $W_c$. In this regime, the $\beta$ dependence of the tails also imparts its influence on the bulk of the distributions.

As explained above, $\psi(t)$ is essentially the Laplace transform of the probability distribution of the imaginary part of the Green's function (see Eq.~\eqref{eq:Phatgs}), with the change of variable $t = \ln s$. Hence, the region around the kink corresponds to the values of $\hat{m}$ of order $1$, while the exponential tails at $t \ll -1$ of $\psi(t)$, Eq.~\eqref{eq:psi_tail}, correspond to the power-law tails of $Q(0,\hat{m})$ for $\hat{m} \gg 1$ with exponent $1+\beta$. Drawing inspiration from our examination of the NL$\sigma$M, below we endeavored to apply a similar scaling analysis to the Anderson model. 
In order to do so, we define $\delta \tilde{Q}(\hat{m};W)$ as the difference between the order parameter distribution function $Q(0,\hat{m})/\avg{\rho}$ found at disorder $W>W_c$ and right at the critical point $W=W_c$:
\begin{equation} \label{eq:deltaQ}
    \delta \tilde{Q}(\hat{m};W) \equiv  \frac{Q(0,\hat{m};W)}{\avg{\rho(W)}} - \frac{Q(0,\hat{m};W_c)}{\avg{\rho(W_c)}}\, ,
\end{equation}
in terms of which one has:
\begin{equation} \label{eq:deltaI}
    I_2(W) - I_2^{(c)} = \int \frac{\delta \tilde{Q}(\hat{m};W)}{\hat{m}} \, \de \hat{m} \, .
\end{equation}
In Fig.~\ref{fig:deltaQ} we show that, when divided by the square root distance from the critical point $(\delta W)^{1/2} = \sqrt{W-W_c}$, the $\delta\tilde{Q} (\hat{m};W)$'s computed for different disorder levels tend to collapse on the same scaling function when $W$ approaches $W_c$. The right panel of Fig.\ref{fig:deltaQ} highlights this data collapse particularly in the region $\hat{m} \gtrsim 1$, which gives the dominant contribution to the integral~\eqref{eq:deltaI}. This implies that the square root singularity of the IPR observed in Fig.~\ref{fig:IPRc} is due to the square root dependence of $\delta \tilde{Q}$ in the bulk, and in particular at small $\hat{m}$. In a specular way, the collapse implies that at large $\hat{m}$ and near $W_c$, the tails of the order parameter distribution function behave as:
\begin{equation}
Q(0,\hat{m}) \simeq \frac{c_c + c_0 \sqrt{W-W_c}}{\hat{m}^{\frac{1}{2} + \sqrt{\frac{c_1}{c_2}} \sqrt{W - W_c} } } \, ,
\end{equation}
(which is the analog of Eq.~\eqref{eq:deltapsitails}) where $c_1$ and $c_2$ are given in Eq.~\eqref{eq:coefficients} for $k=2$~\cite{tikhonov2019critical}. The mechanism by which the square root dependence of the {\it prefactor} of the tails is directly inherited from the square root dependence of the {\it exponent} of the tails is not obvious, and is certainly an interesting question for future investigations.

\section{Conclusions} \label{sec:conclusions}
In this paper we have analyzed the localized phase of the Anderson model on the infinite BL. We have put forward an improved population dynamics scheme to compute the moments of the imaginary part of the Green's function 
directly in the limit $\eta=0$ with unprecedented accuracy even very close to the critical point. This approach allows one to validate the critical behavior predicted by the supersymmetric analysis~\cite{efetov1985anderson,efetov1987density,efetov1987anderson,zirnbauer1986localization,zirnbauer1986anderson,verbaarschot1988graded,mirlin1991localization,mirlin1991universality,fyodorov1991localization,fyodorov1992novel,mirlin1994statistical,tikhonov2019statistics,tikhonov2019critical} with very high accuracy. It also unveils a remarkable feature that has not been reported in the previous literature: The finite jump of the IPR at the transition is followed by a square root singularity, whose existence is also confirmed by the analysis of the effective NL$\sigma$M formulation of the problem on the BL.

It would be interesting to interpret this result in terms of the geometric structure of localized eigenstates on the BL, and understand whether the singular behavior of the IPR is related to the one of the transverse localization length which controls the exponential decay of the wave-functions on typical branches~\cite{garcia2020two,garcia2022critical}.

Ultimately, delving into the loop corrections to the BL solution of AL and broadening the analysis of Ref.~\cite{baroni2024corrections} to encompass the insulating phase presents a highly intriguing prospect. Given that a comprehensive understanding of the loop corrections hinges on mastering very precisely the BL solution, the current investigation serves as a pivotal stride forward, laying the foundation for further exploration in this direction.

\begin{acknowledgments}
We warmly thank Y.~Fyodorov, G. Lemarié and A. D. Mirlin  for illuminating discussions.
\end{acknowledgments}

\begin{appendix}
\begin{widetext}
\section{Stability of the linearized equation} \label{app:Kernel}

The linearized cavity equations~\eqref{eq:real} and~\eqref{eq:imag} must be interpreted as a self-consistent integral equation for the probability distribution $P(g,\hat{g})$:
\[
P(g,\hat{g}) = \int \de \epsilon p(\epsilon) \prod_{i=1}^k \left[ \de g_i \, \de \hat{g}_i P(g,\hat{g}) \right] \delta \left(g - \frac{1}{ \epsilon - \sum_{i=1}^k \, {g}_{i} } \right)
\delta \left(\hat{g} - g^2 \left( 1 + \sum_{i=1}^k \hat{g}_i \right) \right)\, .
\]
This equation is more conveniently written performing the Laplace transform with respect to $\hat{g}$~\cite{anderson1973selfconsistent} (note $\hat{g}$ takes only strictly positive values):
\begin{equation} \label{eq:Phatgs}
\hat{P}(g,s)=\int \de \epsilon p(\epsilon) \prod_i^k \left[ \hat{P}(g_i,s g^2)\, \de g_i\right]  \delta \left(g - \frac{1}{ \epsilon - \sum_{i=1}^k {g}_{i} } \right) e^{- s g^2} \, . 
\end{equation}
Following Ref.~\cite{anderson1973selfconsistent} we identify the localization transition as the point where the above equation ceases to have a solution. To do so we focus on the  region $s \ll 1$, corresponding to the tail of the probability of the imaginary part, $\hat{g} \gg 1$. At small values of $s$ we assume that:
\begin{equation} 
\hat{P}(g,s) \approx P_0(g) + f (g) \, s^{\beta} \, ,
\label{asymptotic}
\end{equation}
where $P_0(g) \equiv \hat{P}(g,0)$ is the probability distribution of the real part of the Green's function, corresponding to the solution of Eq.~\eqref{eq:Phatgs} with $\eta=0$. Note that the ansatz~\eqref{asymptotic} corresponds to Eq.~\eqref{eq:ansatz} of the main text. Plugging the above small-$s$ form into the equation~\eqref{eq:Phatgs} we obtain the following linear equation for the function $f (g)$:
\begin{equation} \label{eq:fg}
f (g)=k \int \de \epsilon p(\epsilon) \prod_{i=2}^{k} \left[ P_0(g_i)\, \de g_i \right] \delta \left(g - \frac{1}{\epsilon - \sum_{i=1}^k {g}_{i} } \right) |g|^{2\beta} f (g_1) \, \de g_1  
\end{equation}
We now introduce the probability distribution of the sum of the real part of $k-1$ cavity Green's functions:
\begin{equation} \label{eq:Ptilde}
\tilde{P} (\tilde{g}) \equiv \int \prod_{i=1}^{k-1} \left[ P_0 (g_i) \, \de g_i \right] \delta \left( \tilde{g} - \sum_{i=1}^{k-1} g_i \right) \, .
\end{equation}
(Note that for $k=2$ one has that $\tilde{P} (x) = P_0 (x)$~\cite{tikhonov2019critical}.) Inserting this identity into Eq.~\eqref{eq:fg} we obtain:
\begin{equation} \label{eq:fg1}
f (g)=k \int \de \epsilon p(\epsilon) \, \de \tilde{g} \tilde{P} ( \tilde{g} ) \, \delta \left(g - \frac{1}{\epsilon - g_1 - \tilde{g} } \right) |g|^{2\beta} f (g_1) \, \de g_1 \, ,
\end{equation}
which coincides with Eqs.~\eqref{eq:LIO} and~\eqref{eq:Kernel} of the main text.

The condition that the above homogeneous equation admits a solution fixes the value of $\beta$. This is only possible if the largest eigenvalue $\lambda_\beta$ of the integral operator is smaller than 1. The critical point is identified by the point where no solution exists. It is easy to check that the probability distribution of the real part of the Green's function $P_0(g)$ is an eigenvector of the integral operator for $\beta=0$, corresponding to the largest eigenvalue $k$ (see Fig.~\ref{fig:kernel}).  As explained in the main text, it can be shown that $\beta=1/2$ at the critical point. In fact, since the integral operator above is non-symmetric, for each eigenvalue, there will be a right and left eigenvector. After inegrating over the $\delta$-function, using the fact that $\delta (w(\tilde{g})) = \delta(\tilde{g}_0)/|w^\prime(\tilde{g}_0)|$, with $\tilde{g}_0 = \epsilon - g_1 - 1/g$ and $|w^\prime(\tilde{g}_0)| = g^2$, one gets:
\[
\begin{aligned}
    \lambda_\beta \, \psi_\beta (g) & =  k \, |g|^{2(\beta-1)}  \int \de \epsilon p(\epsilon) \,  \tilde{P} \left( \epsilon -g_1 - \frac{1}{g} \right) \psi_\beta (g_1) \, \de g_1 \, , \\
     \lambda_\beta \, \phi_\beta (g_1) & = k \int \de \epsilon p(\epsilon) \,  \tilde{P} \left( \epsilon -g_1 - \frac{1}{g} \right) |g|^{2(\beta-1)} \phi_\beta (g) \, \de g \, .
\end{aligned}
\]
From the second equation, defining $\psi_{1-\beta} (g_1) = |g_1|^{-2 \beta} \phi(1/g_1)$ and changing variable $g \to 1/g$ in the left hand side, one sees that $\psi_{1-\beta} (g_1)$ is a right eigenvector of the integral operator~\eqref{eq:Kernel} for $\beta \to 1 - \beta$ with the same eigenvalue $\lambda_\beta$. Hence the spectrum of~\eqref{eq:Kernel}, and in particular its largest eigenvalue, must be symmetric around $\beta=1/2$, as schematically illustrated in Fig.~\ref{fig:kernel}.

\section{Algorithm to evaluate Eq.~\eqref{eq:IpQ}} \label{sec:algoIp}

The algorithm implemented to evaluate Eq.~\eqref{eq:IpQ} and compute $Q(0,\hat{m})$ is schematically summarized as follows:
\begin{tabbing} xx \= xx \= xx \= xx \= xx \= xx \= xx \= xxxxx \kill
 {\bf Algorithm} computing $I_p$ and $Q(0,\hat{m})$\\
 {\bf begin}\\
 \> Initialize population of $\Omega$ elements $(g_\alpha, \hat{g}_\alpha)_{\alpha = 1, \ldots, \Omega}$\\
 \> Iterate population using Eqs.~\eqref{eq:real} and~\eqref{eq:imag}
     until convergence to a stationary distribution\\
 \> {\bf begin}\\
\>\> $I_p = 0$; $Q(0,\hat{m})=0$ \\ 
 \>\> {\bf for} r=1 to $N_{\rm avg}$ {\bf do}\\
 \>\>\> $a=0$; $b=0$ \\
 \>\>\> {\bf for} i=1 to $N_{\rm est}$ {\bf do}\\
 \>\>\>\> Sample $k+1$ elements from the pool $(g_{\alpha_j}, \hat{g}_{\alpha_j})$, $j=1,\ldots,k+1$\\ 
 \>\>\>\> Extract a random energy $\epsilon$ from the box distribution of width $W$\\
 \>\>\>\>  Compute $S = \sum_{j=1}^{k+1} g_{\alpha_j}$\\
 \>\>\>\> {\bf if} $|S| < W/2$ {\bf then}\\
   \>\>\>\>\> Compute $\hat{m}$ from Eq.~\eqref{eq:mii1}: $\hat{m} = 1 + \sum_{j=1}^{k+1} \hat{g}_{\alpha_j}$\\
 \>\>\>\>\> $a = a+\hat{m}^{1-p}/W$\\
  \>\>\>\>\> $b=b+1/W$ \\
 \>\>\>\>\> Add $\hat{m}$ to $Q(0,\hat{m})$ \\
 \>\>\>\> {\bf end if}\\
 \>\>\> {\bf end for}\\
 \>\>\> $a=a/N_{\rm est}$ \\
 \>\>\> $b=b/N_{\rm est}$\\
 \>\>\> $I_p= I_p+ a/b$\\
 \>\>\> Renew all the elements of the population \\
 \>\> {\bf end for}\\
 \>\> $I_p= I_p/N_{\rm avg}$\\
 \>\> Normalize $Q(0,\hat{m})$ \\ 
 \> {\bf end}\\
 {\bf end}\\
\end{tabbing}
\end{widetext}

\section{Exact diagonalizations of the Anderson model on RRGs of $N$ nodes} \label{app:ED}

We preform exact diagonalizations of the Anderson tight-binding model~\eqref{eq:H} on {\it finite} Bethe lattices of fixed connectivity $k+1=3$. Finite BLs are in fact random-regular graphs of $N$ nodes, a class of random lattices that have locally a tree-like structure but do not have boundaries. More precisely, a $(k+1)$-RRG is a lattice chosen uniformly at random among all possible graphs of $N$ vertices where each of the sites has fixed degree $k+1$. The properties of such random graphs have been extensively studied (see Ref.~\cite{wormald1999models} for a review). A RRG can be essentially thought as a finite portion of a tree wrapped onto itself. It is known in particular that for large number of vertices any finite portion of  a RRG is a tree with a probability going to one as $N \to \infty$: RRGs have loops of all size but  short loops are rare and their typical length is of order $\ln N/\ln k$~\cite{wormald1999models}.

Thanks to the sparse nature of the graph, exact diagonalizations can be efficiently performed using the Arnoldi method, which provides a few eigenvalues and eigenvectors around $E=0$. In practice we consider the $64$ nearest eigenstates to zero energy. When comparing the results obtained from exact diagonalizations 
with the analytic predictions obtained at $E=0$, a suitable approach is taken to minimize corrections arising from the small deviation of eigenvalues from precisely zero energy. This is achieved through the following procedure: We start by noticing that the eigenvectors of $H$ are also eigenvectors of $H + \gamma \mathbb{I}$ with all eigenvalues shifted by $\gamma$, \ie~$E_\alpha \to E_\alpha + \gamma$. Thus an eigenvector of $H$ of energy $E_\alpha$ is an eigenvector of zero energy of an Anderson model~\eqref{eq:H} with all random energies shifted by $-E_\alpha$ (note that since we consider only a finite number of eigenvectors, the $E_\alpha$'s are of order $1/N$, and thus only a few random energies $\epsilon_i - E_\alpha$ will fall outside the box of width $W$ after the shift). Since $\Tr H = \sum_i \epsilon_i$ converges to a normal distribution with zero mean and variance $N W^2/12$, shifting the trace of $H$ by $N E_\alpha$ must be reweighted by a factor $e^{- 6 N E_\alpha^2/W^2}$. As a result, to obtain the averages at zero energy of a generic observable which depends on the wave-functions' amplitudes, we use the following expression:
\[
\avg{O(\{ \psi_\alpha (i) \} )} = \frac{\sum_\alpha e^{- 6 N E_\alpha^2/W^2} O(\{ \psi_\alpha (i) \}) } {\sum_\alpha e^{- 6 N E_\alpha^2/W^2}} \, .
\]

\end{appendix}

\bibliography{references}

\end{document}